\documentclass[aps,nofootinbib]{revtex4}
\usepackage{natbib}
\usepackage{graphicx}
\usepackage{amssymb}
\usepackage{subfigure}
\usepackage{stackrel}
\usepackage{amsfonts}
\usepackage{bm}
\usepackage{amsmath}
\usepackage{xcolor}
\usepackage[normalem]{ulem}
\usepackage{hyperref}
\usepackage{verbatim}

\begin{document}

\title{Asymptotic safe nonassociative quantum gravity with star R-flux
products, Goroff-Sagnotti counter-terms, and geometric flows}
\date{revised: September 9, 2024}
\author{Sergiu I. Vacaru}
\email{ sergiu.vacaru@fulbrightmail.org ; sergiu.vacaru@gmail.com}

\begin{abstract}
Nonassociative modifications of general relativity, GR, defined by star
products with R-flux deformations in string gravity consist an important
subclass of modified gravity theories, MGTs. A longstanding criticism for
elaborating quantum gravity, QG, argue that the asymptotic safety does not
survive once certain perturbative terms (in general, nonassociative and
noncommutative) are included in the projection space. The goal of this work
is to prove that a generalized asymptotic safety scenario allows us to
formulate physically viable nonassociative QG theories using effective
models defined by generic off-diagonal solutions and nonlinear symmetries in
nonassociative geometric flow and gravity theories. We elaborate on a new
nonholonomic functional renormalization techniques with parametric
renormalization group, RG, flow equations for effective actions supplemented
by certain canonical two-loop counter-terms. The geometric constructions and
quantum deformations are performed for nonassociative phase spaces modelled
as R-flux deformed cotangent Lorentz bundles. Our results prove that
theories involving nonassociative modifications of GR can be well defined
both as classical nonassociative MGTs and QG models. Such theories are
characterized by generalized G. Perelman thermodynamic variables which are
computed for certain examples of nonassociative geometric and RG flows.

\vspace{1mm} \noindent\textbf{Keywords}: Nonassociative quantum gravity;
string gravity and R-flux deformations; asymptotic safe gravity theories;
off-diagonal solutions; nonassociative geometric flow thermodynamics.
\end{abstract}
\address{{\small \textit{Taras Shevchenko National University of Kyiv,
Astronomical Observatory, Kyiv, Ukraine }} }
\maketitle

%\tableofcontents

%%%

\section{Introduction}

\label{sec1}Nonassociative and noncommutative models of gravity with twisted
star product, $\star $, \cite{drinf89} can be elaborated as effective
modified gravity theories, MGTs, generalizing Einstein gravity, i.e. the
general relativity, GR, theory. Such nonassociative MGTs are expected to
arise in low-energy limits of string theories \cite%
{luest10,blumenhagen10,condeescu13,blumenhagen13}. Nonassociative gravity
was formulated in self-consistent forms up to the definition of
nonassociative vacuum Einstein equations for phase spaces defined as $\star $%
-product deformations of cotangent Lorentz bundles in \cite%
{blumenhagen16,aschieri17}. A series of recent articles \cite%
{partner02,partner04,partner05,partner06} was devoted to performing a
research program on nonassociative geometry, physics and quantum geometric
and information theory. In that framework, an advanced geometric techniques
(the so-called anholonomic frame and connection deformation method, AFCDM)
was developed with the aim to constructing generic off-diagonal exact and
parametric solutions in nonassociative and noncommutative geometric flow and
gravity theories. Such new types of solutions may describe nonassociative
black hole, BH, and wormhole, WH, configurations, or define cosmological
scenarios encoding string R-flux contributions. The nonassociative geometric
models involve nonholonomic non-Riemannian geometric objects, with
nonsymmetric and symmetric metrics and (non) linear connections, when the
coefficients may depend on all spacetime and momentum-like variables. In
such nonassociative theories, physically important systems of nonlinear
partial differential equations, PDEs, can be derived in abstract geometric
form but not in a general variational form. This is not possible for general
twist nonassociative star products. Nevertheless, self-consistent
variational methods (with effective Lagrange and Hamilton densities and, for
our purposes, nonassociative modified Perelman's F- and W-functionals \cite%
{perelman1}) can be elaborated for parametric decompositions on the Planck
and string constants, $\hbar $ and $\kappa =\mathit{\ell }_{s}^{3}/6\hbar $,
when the string fundamental length is taken as $\mathit{\ell }_{s}$ (see
details in \cite{aschieri17,partner04,partner05,partner06}).

String theory was elaborated as an unified approach to fundamental physics
(see the monographs \cite{string1,string2,string3} as comprehensive reviews)
with an aim to incorporate quantum gravity, QG, classical and quantum field
theories, QFTs, and which in semi-classical associative and commutative
limits results in GR defined by the Einstein-Hilbert action. Various models
with non-Riemanian geometric and physical objects and MGTs where also
considered. The main results of \cite%
{blumenhagen16,aschieri17,partner04,partner05,partner06} state that twisted
star products determined by nontrivial R-flux contributions modify GR both
at classical and quantum levels and result in new types of generic
off-diagonal interactions and nonlinear symmetries of nonassociative phase
spaces. A central puzzle for constructing a consistent quantum theory of
gravity and matter fields encoding nonassociative and noncommutative data is
its perturbative non-renormalizability of projections to Lorentz spacetime
manifolds and related effective theories. For Einstein's gravity, the
gravitational two-loop counter-term is asymptotically safe \cite%
{weinberg79,niedermaier06,gies16}, see \cite%
{bonanno11,dietz13,benedetti13,ohta15} and references therein for recent
results on asymptotic properties of MGTs. In this context, an important
issue is to investigate if star products and off-diagonal interactions
preserve the property of asymptotic safe when the QG models are elaborated
for nonassociative phase spaces.

Any consistent QG theory allowing a semi-classical extension as a low-energy
effective theory for quantized gravitons, with S-matrix elements determined
by the Einstein-Hilbert action related to an expansion in terms of the
Newton's constant, gives rise to divergences at two-loop order \cite%
{goroff1,goroff2,ven92}. This is described by a nontrivial Goroff-Sagnotti
counter-term which is a cubic operator, $C^{3},$ determined by contractions
of respective pairs of indices in the Weyl tensor $C_{ijkl}.$\footnote{%
We follow the conventions and notations from \cite%
{partner04,partner05,partner06} when $i,j,...=1,2,3,4$ for a
four-dimensional, 4-d, Lorentz spacetime manifold $V$ of signature $(+++-),$
or $(++-+);$ in the next footnotes and sections, we shall provide necessary
formulas for $\star $-deformations of geometric objects on nonassociative
8-d phase spaces modelled on a cotangent Lorentz bundle $T^{\ast }V$.} To
analyze the perspective of including nonassociative gravity theories of type 
\cite{blumenhagen16,aschieri17,partner04,partner05,partner06} in a
self-consistent form in QG we have to clarify the fate of the divergencies
related to nonassociative $\star $-product deformations of fundamental
geometric and physical objects, for instance, of the Riemannian, Ricci and
Weyl tensors and related Goroff-Sagnotti counter-term (all computed at least
as parametric deformations on the first order on $\hbar $ and $\kappa $).
For GR, the Wilsonian viewpoint of renormalization offers a solution to this
puzzle. This is possible even for the systems with breakdown of perturbative
quantization, where a well-controlled ultraviolet, UV, limit can be found if
a fixed point exists for the renormailizaton group, RG, equations \cite%
{wilson73,weinberg79,niedermaier06,gies16}. Such results were generalized
for various associative and (non) commutative MGTs \cite%
{bonanno11,dietz13,benedetti13,ohta15}.

The main goal of this work is to show that the Wilsonian approach to
renormalization can be generalized for the nonassociative star product
gravity with modifications determined by R-flux deformations in string
theory. We prove that higher-dimensional operators on phase spaces decouple
from the low-energy physics in some forms which are proportional to an
inverse power of a high scale $\Lambda ^{sc}.$ The main assumptions are
those that respective operators do not acquire large anomalous dimensions,
when the spacetime RG fixed points are taken as in GR and the nonholonomic
structure on a corresponding phase space (modelled on cotangent Lorentz
manifolds, in general, with R-flux deformations) can be fixed in a form when
the canonical Weyl tensor vanish at least up to quadratic orders on $\hbar $
and $\kappa .$ To perform such constructions is not just a technical problem
of finding off-diagonal solutions of nonlinear PDEs because of existence of
new nonassociative and off-diagonal symmetries at a more fundamental level.
We show how keeping the Wilson viewpoint with RG flow equations for the
Einstein theory it is possible to generalized the approach to nonassociative
geometric flows \cite{partner04,partner05,partner06} with respective
projections to Riemannian \cite{perelman1} and pseudo-Riemanian relativistic
flows. This is crucial for elaborating a self-consistent approach to
classical MGTs and QG with R-flux contributions from string theory.

The paper is structured as follows. In section \ref{sec02}, we outline the
main concepts and necessary formulas for the nonassociative gravity and
geometric flow theory with star product R-flux deformations. The AFCDM for
constructing exact and off-diagonal solutions in such MGTs and their
nonlinear symmetries are explained in section \ref{sec03}. Section \ref%
{sec04} is devoted to the nonholonomic functional renormalization procedure
performed on nonassociative phase spaces. We prove that beta functions and
RG flow equations resulting in fixed points for the asymptotic safe QG model
of the Einstein-Hilbert action can be included self-consistently into the
theory of nonassociative geometric flows with flow evolution equations with
general decoupling and integration properties. Such exact or parametric
solutions defining nonholonomic Ricci solitons, which are equivalent to
modified Einstein equations, are characterized by generalized G. Perelman
thermodynamic variables. We conclude the paper in section \ref{sec05}.

\section{Nonassociative star product deformations of GR and geometric flows}

\label{sec02}Nonassociative star product R-flux modifications of gravitational field equations in GR and MGTs can be considered  certain examples of nonassociative Ricci solitons for nonassociative geometric flows
on a phase space $\ ^{\shortmid }\mathcal{M}:=T^{\ast }\mathbf{V}$ defined as a cotangent bundle to a Lorentz manifold $V$ where studied in a series of our partner works \cite{partner02,partner04,partner05,partner06}. In this section, we summarize the necessary results  and methods. 

Let us outline the conventions we follow in this paper.  By definition $\mathcal{M}:=T\mathbf{V}$ is a phase space parameterized locally by base Lorentz spacetime coordinates, $x=\{x^{i}\},$ and velocity type fiber coordinates $y=\{y^{a}\}.$ The dual tangent bundle, i.e. cotangent bundle, $T^{\ast }\mathbf{V,}$ is parameterized with dual momentum type cofiber coordinates   $p=\{p_{a}\}.$ The abstract geometric formalism for Lorentz manifolds from \cite{misner} can be extended both on  $\mathcal{M}$ and 
$\ ^{\shortmid }\mathcal{M}$ by omitting indices and coordinates. It is convenient to use a left-label 
$\ "^{\shortmid }"$  for spaces and geometric objects modeled on  $\ ^{\shortmid }\mathcal{M}$ involving
dependence on momentum like coordinates $p_{a}.$ The total space coordinates will
be denoted respectively   $u=(x,y),$ or $u^{\alpha }=(x^{i},y^{a}),$ on  $\mathcal{M};$ and $\ ^{\shortmid }u=(x,p),$ or $\ ^{\shortmid }u^{\alpha }=(x^{i},p_{a}),$ on  $\ ^{\shortmid }\mathcal{M},$  when the
indices $\alpha ,\beta ,\gamma ,\delta ,..=1,2,...8,$ for $\dim \ ^{\shortmid }\mathcal{M}=\dim \ ^{\shortmid }\mathcal{M}=8,$ In this work, we shall perform our geometric constructions on 
$\ ^{\shortmid }\mathcal{M}$ because the definition of nonassociative  twisted product R-flux
deformations of GR and MGTs \cite{blumenhagen16,aschieri17,partner02,partner04,partner05,partner06} were performed on $\star $-deformed cotangent Lorentz bundles, see details in next subsection.   

The geometric constructions were performed in a canonical nonholonomic form
involving nonlinear connection, N-connection, structures. Such a
N-connection can be defined by a Whitney sum, $\ ^{\shortmid }\mathbf{N:}\
TT^{\ast }\mathbf{V}=hT^{\ast }\mathbf{V\oplus }vT^{\ast }\mathbf{V}$, with
conventional horizontal, $h$, and vertical, $v$, projections for a
nonholonomic 4+4 splitting. Using oriented nonholonomic dyadic shell adapted
variables for a conventional (2+2)+(2+2) splitting and distortions of the
Levi-Civita, LC, connection to a necessary type of auxiliary connection, we
proved general and very important decoupling and integration properties of
nonassociative modified geometric flow and gravitational field equations.
Any phase space $\ ^{\shortmid }\mathcal{M}$ can be enabled additionally
with a (3+1)+(3+1) splitting which is necessary for formulating phase space
generalizations of the Arnowitt-Deser-Misner, ADM, methods and abstract
geometric formalism for GR \cite{misner}, which are necessary for
constructing relativistic and nonassociative versions of the G. Perelman
thermodynamics \cite{perelman1}.

\subsection{Nonassociative phase space geometry}

We can define a (twist) nonassociative star product in such a nonholonomic
dyadic form: 
\begin{eqnarray}
f\star _{s}q &:=&\cdot \lbrack \exp (-\frac{1}{2}i\hbar (\ ^{\shortmid }%
\mathbf{e}_{i_{s}}\otimes \ ^{\shortmid }e^{i_{s}}-\ ^{\shortmid
}e^{i_{s}}\otimes \ ^{\shortmid }\mathbf{e}_{i_{s}})+\frac{i\mathit{\ell }%
_{s}^{4}}{12\hbar }R^{i_{s}j_{s}a_{s}}(p_{a_{s}}\ ^{\shortmid }\mathbf{e}%
_{i_{s}}\otimes \ ^{\shortmid }\mathbf{e}_{j_{a}}-\ ^{\shortmid }\mathbf{e}%
_{j_{s}}\otimes p_{a_{s}}\ ^{\shortmid }\mathbf{e}_{i_{s}})]f\otimes q 
\notag \\
&=&f\cdot q-\frac{i}{2}\hbar \lbrack (\ ^{\shortmid }\mathbf{e}_{i_{s}}f)(\
^{\shortmid }e^{i_{s}}q)-(\ ^{\shortmid }e^{i_{s}}f)(\ ^{\shortmid }\mathbf{e%
}_{i_{s}}q)]+\frac{i\mathit{\ell }_{s}^{4}}{6\hbar }%
R^{i_{s}j_{s}a_{s}}p_{a_{s}}(\ ^{\shortmid }\mathbf{e}_{i_{s}}f)(\
^{\shortmid }\mathbf{e}_{j_{s}}q)+\ldots .  \label{starpn}
\end{eqnarray}%
In these formulas, the phase space coordinates $\alpha _{s}=(i_{s},a_{s})$
contain an additional label $s=1,2,3,4$ for a corresponding 2-d oriented
shell (for instance, $i_{3}=1,2,3,4$ and $a_{3}=5,6;i_{4}=1,2,...,5,6$ and $%
a_{4}=7,8$). We omit the $s$-label for a nonholonomic (4+4)-splitting and
write $\star _{N},$ when $\ ^{\shortmid }\mathbf{e}_{\alpha }=(\ \mathbf{e}%
_{i}=\partial /\partial x^{i}-\ ^{\shortmid }N_{ai}(x^{j},p_{b})\
^{\shortmid }\partial ^{a},\ ^{\shortmid }e^{a}=$ $\ ^{\shortmid }\partial
^{a}=$ $\partial /\partial p_{a},$ for $i,j,..,=1,2,3,4$ and local
coefficients of$\ ^{\shortmid }\mathbf{N}=\{\ ^{\shortmid }N_{ai}\}.$%
\footnote{%
We use boldface symbols for geometric objects (in brief, distinguished, or
d-objects; or shell, s-objects which can be adapted to a respective
N-connection structure).} In dual form, we can define and compute the
N-adapted components of respective dual bases, for instance of $\
^{\shortmid }\mathbf{e}^{\alpha }=(dx^{i},\ ^{\shortmid }\mathbf{e}%
_{a}=...). $ The noncommutative structure of a star product (\ref{starpn})
is defined by terms proportional to the Planck constant $\hbar $ and the
nonassociative structure is determined by some R-flux coefficients in a
string theory, $R^{i_{s}j_{s}a_{s}},$ with a corresponding string length $%
\mathit{\ell }_{s}$ constant. Using holonomic frames $\ ^{\shortmid }\mathbf{%
e}_{\alpha }\rightarrow \ ^{\shortmid}\partial _{\alpha }$, we obtain the
formulas for the star product defined in \cite{blumenhagen16,aschieri17}.
The holonomic variant allows us to define and compute nonassociative Ricci
tensors, and formulate nonassociative vacuum gravitational equations, but
does not allow a general decoupling and integration in certain general forms
of physically important systems of nonlinear PDEs. In another turn, we
proved \cite{partner02,partner04,partner05,partner06} that a formulation of
nonassociative geometric flow theory using the definition (\ref{starpn}) and
respective distortions of generalized connections allows to develop the
AFCDM and apply it for decoupling and solving vacuum and nonvacuum
nonassociative modified Einstein and R. Hamilton equations \cite%
{hamilton82,perelman1}.

A nonholonomic $\star _{N}$-product structure transforms a phase space into
a nonassociative one, $\ ^{\shortmid }\mathcal{M}\rightarrow \
_{s}^{\shortmid }\mathcal{M}^{\star } \simeq \ ^{\shortmid }\mathcal{M}%
^{\star },$ and provide us with an analytic method for computing star
product deformations of fundamental (commutative) geometric objects into
nonassociative ones.\footnote{%
In our approach, an abstract label $s$, or $\star $, can be used on
convenience (up, or low; on the left/ right of some symbols for geometric
objects).} For instance, a metric structure transforms as $\ ^{\shortmid }%
\mathbf{g}=\{\ ^{\shortmid }\mathbf{g}_{\alpha \beta }(\ ^{\shortmid
}u)\}\rightarrow \ ^{\shortmid }\mathbf{g}^{\star }= \{\ ^{\shortmid }%
\mathbf{g}_{\alpha \beta }^{\star }(\ ^{\shortmid }u)\}$ when $\ ^{\shortmid
}\mathbf{g}^{\star }$ may be nonsymmetric and with a sophisticate rule for
computing the inverse $\star $-metric $\ ^{\shortmid }\mathbf{g}_{\star
}^{\alpha \beta }$. Another important example is that of star product
deformations of an arbitrary affine connection structure, $\ ^{\shortmid
}D=\{\ ^{\shortmid }\Gamma _{\ \beta \gamma }^{\alpha }\}\rightarrow \
^{\shortmid }D^{\star }=\{\ ^{\shortmid}\Gamma _{\ \star \beta \gamma
}^{\alpha }\},$ when the nonassociative geometric constructions may be also
metric noncompatible. In \cite{blumenhagen16,aschieri17}, there were
considered $\star $-analogs of the Levi-Civita, LC, connection (which, by
definition, is metric compatible and with zero torsion), $\ ^{\shortmid
}\nabla \rightarrow \ ^{\shortmid}\nabla ^{\star },$ and respective Ricci
tensor, $\ ^{\shortmid }Ric[\ ^{\shortmid }\nabla ]\rightarrow \ ^{\shortmid
}Ric^{\star }[\ ^{\shortmid }\nabla ^{\star }],$ for instance, with
nonsymmetric $\ ^{\shortmid }Ric^{\star }=\{\ ^{\shortmid }R_{\alpha \beta
}^{^{\star }}\}$ for a nontrivial nonholonomic structure. A tedious
geometric and analytic procedure allows us to define and compute star
product deformations the Ricci scalar $\ ^{\shortmid }Rs[\ ^{\shortmid
}\nabla ]\rightarrow \ ^{\shortmid }Rs^{\star }[\ ^{\shortmid }\nabla
^{\star }]$; with $\ ^{\shortmid }Rs^{\star }:=\ ^{\shortmid }\mathbf{g}%
_{\star }^{\alpha \beta }\ ^{\shortmid }R_{\alpha \beta }^{^{\star }},$ when
the Einstein summation rule is applied. The nonholonomic N- or s-adapted
calculus considered in \cite{partner02,partner04,partner05,partner06} can be
applied to compute the geometric objects in corresponding abstract and
adapted component forms which is stated by "boldface" symbols, for instance, 
$\ ^{\shortmid }\mathbf{D}\rightarrow \ ^{\shortmid }\mathbf{D}^{\star }, \
^{\shortmid }\mathbf{R}ic^{\star }=\{\ ^{\shortmid }\mathbf{R}_{\alpha \beta
}^{^{\star }}\}$ etc.

For any geometric data $(\ _{s}^{\shortmid }\mathbf{N,}\ _{s}^{\ \shortmid }%
\mathbf{g}^{\star }),$ we can define a canonical s-connection (in brief, the
term "hat-connection" etc., defining fundamental canonical geometric
s-objects), with respective shell projections, $\ _{s}^{\shortmid }\widehat{%
\mathbf{D}}^{\star }=(h_{1}\ ^{\shortmid }\widehat{\mathbf{D}}^{\star },\
v_{2}\ ^{\shortmid }\widehat{\mathbf{D}}^{\star },\ c_{3}\ ^{\shortmid }%
\widehat{\mathbf{D}}^{\star },\ c_{4}\ ^{\shortmid }\widehat{\mathbf{D}}%
^{\star })=\ ^{\shortmid }\nabla ^{\star }+\ _{s}^{\shortmid }\widehat{%
\mathbf{Z}}^{\star }$, where the canonical distortion s-tensor $\
_{s}^{\shortmid }\widehat{\mathbf{\ Z}}^{\star }[\ _{s}^{\shortmid }\widehat{%
\mathcal{T}}^{\star }[\ _{s}^{\shortmid }\mathbf{N,}\ _{s}^{\shortmid }%
\mathbf{g}^{\star }]]$ is an algebraic functional of the canonical s-torsion 
$\ _{s}^{\shortmid }\widehat{\mathcal{T}}^{\star }.$ Such fundamental
nonassociative geometric s-objects are defined: {\small 
\begin{equation}
(\ _{s}^{\shortmid }\mathbf{g^{\star },\ _{s}^{\shortmid }N})\rightarrow
\left\{ 
\begin{array}{cc}
\ ^{\shortmid }\mathbf{\nabla ^{\star }:}\  & 
\begin{array}{c}
\fbox{\ $\ \ ^{\shortmid }\mathbf{\nabla }^{\star }\ _{s}^{\ \shortmid }%
\mathbf{g}^{\star }=0$;\ $_{\nabla }^{\shortmid }\mathcal{T}^{\star }=0$}%
\mbox{\ star
LC-connection}; \\ 
\end{array}
\\ 
\ _{s}^{\shortmid }\widehat{\mathbf{D}}^{\star }: & \fbox{$%
\begin{array}{c}
\ _{s}^{\shortmid }\widehat{\mathbf{D}}^{\star }\ _{s}^{\ \shortmid }\mathbf{%
g}^{\star }=0;\ h_{1}\ ^{\shortmid }\widehat{\mathcal{T}}^{\star }=0,v_{2}\
^{\shortmid }\widehat{\mathcal{T}}^{\star }=0,c_{3}\ ^{\shortmid }\widehat{%
\mathcal{T}}^{\star }=0,c_{4}\ ^{\shortmid }\widehat{\mathcal{T}}^{\star }=0,
\\ 
h_{1}v_{2}\ ^{\shortmid }\widehat{\mathcal{T}}^{\star }\neq 0,h_{1}c_{s}\
^{\shortmid }\widehat{\mathcal{T}}^{\star }\neq 0,v_{2}c_{s}\ ^{\shortmid }%
\widehat{\mathcal{T}}^{\star }\neq 0,c_{3}c_{4}\ ^{\shortmid }\widehat{%
\mathcal{T}}^{\star }\neq 0,%
\end{array}%
$}\mbox{ canonical  s-connection }.%
\end{array}%
\right.  \label{twoconsstar}
\end{equation}%
} We note that in the definition of linear connection $\ ^{\shortmid }%
\mathbf{\nabla }^{\star }$ and $\ $s-connection $\ _{s}^{\shortmid }\widehat{%
\mathbf{D}}^{\star }$ we use the same metric s-tensor $_{s}^{\ \shortmid }%
\mathbf{g^{\star }.}$ Similar geometric hat objects can be defined by
canonical 4+4 nonholonomic h- and c-splitting when the s-label can be
omitted. The geometric and physical meaning of a hat-torsion $\ ^{\shortmid }%
\widehat{\mathcal{T}}^{\star }=\{\ ^{\shortmid }\widehat{\mathcal{T}}_{\
\star \beta \gamma }^{\alpha }\}$ are different from the torsions
considered, for instance, in the Riemann-Cartan or string gravity theories
where additional sources are considered in the respective field equations. A
canonical s-torsion or d-torsion is completely defined by generic
off-diagonal metric coefficients for a prescribed nonholonomic structure
with nontrivial N-connection \ coefficients. In principle, we can always
extract LC-configurations with $\ ^{\shortmid }\mathbf{\nabla }_{\ \star
\beta \gamma }^{\alpha }=\ _{s}^{\shortmid }\widehat{\mathbf{D}}_{\ \star
\beta \gamma }^{\alpha }$ in a N-adapted frame system by imposing zero
distortion conditions, $\ _{s}^{\shortmid }\widehat{\mathbf{Z}}^{\star }=0,$
or $\ _{s}^{\shortmid }\widehat{\mathbf{T}}^{\star }=0.$ Such nontrivial
solutions of the zero torsion conditions exist for generic off-diagonal
s-metrics as was proven in \cite{partner02,partner04,partner05,partner06}
which allows us to construct generic off-diagonal solutions even in GR.

Using formulas (\ref{starpn}) \ and (\ref{twoconsstar}) and following
standard methods of nonholonomic differential geometry, we can define and
compute star product deformations of fundamental geometric s-objects. Here
we provide the geometric symbols and respective s-adapted labels: $\
_{s}^{\shortmid }\mathcal{T} \rightarrow \ _{s}^{\shortmid }\widehat{%
\mathcal{T}}^{\star }=\{\ ^{\shortmid }\widehat{\mathbf{T}}_{\ \star \beta
_{s}\gamma _{s}}^{\alpha _{s}}\},$ the nonassociative canonical s-torsion ;\ 
$\ _{s}^{\shortmid }\mathcal{R} \rightarrow \ _{s}^{\shortmid }\widehat{%
\mathcal{R}}^{\star }=\{\ ^{\shortmid }\widehat{\mathbf{R}}_{\ \beta
_{s}\gamma _{s}\delta _{s}}^{\star \alpha _{s}}\}$, the nonassociative
canonical Riemannian s-curvature; $\ _{s}^{\shortmid }\mathcal{C}
\rightarrow \ _{s}^{\shortmid }\widehat{\mathcal{C}}^{\star }=\{\
^{\shortmid }\widehat{\mathbf{C}}_{\ \beta _{s}\gamma _{s}\delta
_{s}}^{\star \alpha _{s}}\}$, the nonassociative canonical Weyl (conformal)
s-tensor; $\ _{s}^{\shortmid }\mathcal{R}ic \rightarrow \ _{s}^{\shortmid }%
\widehat{\mathcal{R}}ic^{\star }=\{\ ^{\shortmid }\widehat{\mathbf{R}}_{\
\beta _{s}\gamma _{s}}^{\star }:=\ ^{\shortmid }\widehat{\mathbf{R}}_{\
\beta _{s}\gamma _{s}\alpha _{s}}^{\star \alpha _{s}}\neq \ ^{\shortmid }%
\widehat{\mathbf{R}}_{\ \gamma _{s}\beta _{s}}^{\star }\}$, the
nonassociative canonical Ricci s-tensor; $\ _{s}^{\shortmid }\mathcal{R}sc
\rightarrow \ _{s}^{\shortmid }\widehat{\mathcal{R}}sc^{\star }=\{\
^{\shortmid }\mathbf{g}^{\beta _{s}\gamma _{s}}\ ^{\shortmid }\widehat{%
\mathbf{R}}_{\ \beta _{s}\gamma _{s}}^{\star }\},$ the nonassociative
canonical Riemannian scalar; $\ _{s}^{\shortmid }\mathcal{Q} \rightarrow \
_{s}^{\shortmid }\mathcal{Q}^{\star }=\{\ ^{\shortmid }\widehat{\mathbf{Q}}%
_{\gamma _{s}\alpha _{s}\beta _{s}\ }^{\star }=\ ^{\shortmid }\widehat{%
\mathbf{D}}_{\gamma _{s}}^{\star }\ ^{\shortmid }\mathbf{g}_{\alpha
_{s}\beta _{s}}^{\star }\}=0,$ the nonassociative canonical nonmetricity
s-tensor.

For the purposes of this work, we consider parametric decompositions of the
star canonical d-connection $\ ^{\shortmid }\widehat{\mathbf{D}}^{\star
}=\{\ ^{\shortmid }\widehat{\mathbf{\Gamma }}_{\star \alpha \beta }^{\gamma
}\}$ or s-connection $\ _{s}^{\shortmid }\widehat{\mathbf{D}}^{\star }$ (\ref%
{twoconsstar}), and respective fundamental nonassociative geometric
s-objects, when 
\begin{equation*}
\ ^{\shortmid }\widehat{\mathbf{\Gamma }}_{\star \alpha \beta }^{\gamma }=\
_{[0]}^{\shortmid }\widehat{\mathbf{\Gamma }}_{\star \alpha \beta }^{\gamma
}+i\kappa \ _{[1]}^{\shortmid }\widehat{\mathbf{\Gamma }}_{\star \alpha
\beta }^{\gamma }=\ _{[00]}^{\shortmid }\widehat{\mathbf{\Gamma }}_{\star
\alpha \beta }^{\gamma }+\ _{[01]}^{\shortmid }\widehat{\mathbf{\Gamma }}%
_{\star \alpha \beta }^{\gamma }(\hbar )+\ _{[10]}^{\shortmid }\widehat{%
\mathbf{\Gamma }}_{\star \alpha \beta }^{\gamma }(\kappa )+\
_{[11]}^{\shortmid }\widehat{\mathbf{\Gamma }}_{\star \alpha \beta }^{\gamma
}(\hbar \kappa )+O(\hbar ^{2},\kappa ^{2}...).
\end{equation*}%
This allows us to compute linear parametric decompositions of type 
\begin{equation}
\ ^{\shortmid }\widehat{\mathbf{R}}_{\ \alpha \beta }^{\star }(\tau )=\
^{\shortmid }\widehat{\mathbf{R}}_{\ \alpha \beta }(\tau )+\ ^{\shortmid }%
\widehat{\mathbf{K}}_{\ \alpha \beta }(\tau ,\left\lceil \hbar ,\kappa
\right\rceil )\mbox{ and }^{\shortmid }\widehat{\mathbf{R}}sc^{\star }=\
^{\shortmid }\widehat{\mathbf{R}}sc+\ ^{\shortmid }\widehat{\mathbf{K}}sc,
\label{paramricci}
\end{equation}%
where$\ ^{\shortmid }\widehat{\mathbf{K}}sc:=\ ^{\shortmid }\mathbf{g}%
_{\star }^{\mu \nu }\ ^{\shortmid }\widehat{\mathbf{K}}_{\ \beta \gamma
}\left\lceil \hbar ,\kappa \right\rceil $ contains the terms proportional to 
$\hbar $ and $\kappa .$

\subsection{Nonassociative geometric flows and G. Perelman thermodynamics}

A theory of nonassociative geometric flows on phase spaces $\
_{s}^{\shortmid }\mathcal{M}^{\star }$ was elaborated in \cite%
{partner04,partner05,partner06} considering star product deformations of
relativistic generalizations of the Hamilton-Perelman theory \cite%
{hamilton82,perelman1} for Riemannian geometric flows. The main goal of our
nonassociative models is to elaborate on applications in classical MGTs and
QG of geometric flow evolution scenarios for $\tau$--families of ($\
^{\shortmid }\mathbf{g}^{\star }(\tau )=\{\ ^{\shortmid }\mathbf{g}_{\alpha
\beta }^{\star }(\tau ,\ ^{\shortmid }u)\}, \ _{s}^{\shortmid }\widehat{%
\mathbf{D}}^{\star }(\tau )),$ where $\tau$ is a temperature like parameter, 
$0\leq \tau \leq \tau _{0}.$ For such nonassociative geometric flow
theories, to formulate and prove certain variants of a nonassociative
Thurston-Poincar\'{e} conjecture is a very difficult and undetermined
mathematical task (in principle, we can consider an infinite number of
nonassociative and noncommutative calculi). Nevertheless, such geometric
constructions are possible for certain QFT and QG theories when
nonassociative RG and geometric flow scenarios arise in a natural way from
string theory. We can study physical properties of respective nonassociative
models in the framework of the theory of parametric nonlinear PDEs and
related nonholonomic geometric methods when the AFCDM allows to decouple and
integrate nonassociative off-diagonal modifications of the R. Hamilton and
Ricci soliton equations.

Using the Ricci d-tensor and respective scalar parametric decompositions (%
\ref{paramricci}), we compute parametric R-flux modifications of Perelman's
F- and W-functionals written in canonical d-variables, 
\begin{eqnarray}
\ ^{\shortmid }\widehat{\mathcal{F}}_{\kappa }^{\star }(\tau )
&=&\int_{^{\shortmid }\Xi }(\ ^{\shortmid }\widehat{\mathbf{R}}sc+\
^{\shortmid }\widehat{\mathbf{K}}sc+|\ \ ^{\shortmid }\widehat{\mathbf{D}}\
\ ^{\shortmid }\widehat{f}|^{2})e^{-\ ^{\shortmid }\widehat{f}}\ d\
^{\shortmid }\mathcal{V}ol(\tau ),\mbox{
and }  \label{naffunctpfh} \\
\ \ ^{\shortmid }\widehat{\mathcal{W}}_{\kappa }^{\star }(\tau )
&=&\int_{^{\shortmid }\Xi }\left( 4\pi \tau \right) ^{-4}\ [\tau (\ \
^{\shortmid }\widehat{\mathbf{R}}sc+\ ^{\shortmid }\widehat{\mathbf{K}}sc+|\
^{\shortmid }\widehat{\mathbf{D}}\ \ ^{\shortmid }\widehat{f}|^{2}\ )^{2}+\
^{\shortmid }\widehat{f}-8]e^{-\ ^{\shortmid }\widehat{f}}\ d\ ^{\shortmid }%
\mathcal{V}ol(\tau ).  \notag
\end{eqnarray}%
In (\ref{naffunctpfh}), the normalizing function $\ ^{\shortmid }\widehat{f}%
(\tau )=$ $\ ^{\shortmid }\widehat{f}(\tau ,\ ^{\shortmid }u)$ is re-defined
to include $\left\lceil \hbar ,\kappa \right\rceil $-terms from $\
^{\shortmid }\widehat{\mathbf{D}}^{\star }(\tau )\rightarrow \ ^{\shortmid }%
\widehat{\mathbf{D}}(\tau )$ and other terms including $\kappa $-parametric
decompositions. In principle, we can chose such a normalization function
which will absorb the distortion terms when $\ ^{\shortmid }\widehat{\mathbf{%
D}}\rightarrow \ ^{\shortmid }\mathbf{\nabla }$ for a nonassociative model
with LC-connections on $\ ^{\shortmid }\mathcal{M}^{\star }$ but using hat
variables we results in systems of nonlinear PDEs with certain general
decoupling and integration properties. Certain closed integration volumes on
the commutative associated phase space $\ ^{\shortmid }\mathcal{M}$ are
denoted $^{\shortmid }\Xi ,$ when the volume form in the $[0]$-approximation
is computed $d\ \ ^{\shortmid }\mathcal{V}ol(\tau )=\sqrt{|\ \ ^{\shortmid }%
\mathbf{g}_{\alpha \beta }\ (\tau )|}\ \delta ^{8}\ ^{\shortmid }u^{\gamma
_{s}}(\tau ),$ where the s-differentials $\delta ^{8}\ ^{\shortmid
}u^{\gamma _{s}}$ are linear on $\ ^{\shortmid }N_{\ i_{s}a_{s}}\ (\tau )$
as in $\ ^{\shortmid }\mathbf{e}_{i_{s}}(\tau ).$

We can fix the normalization function and adapt the nonholonomic structure
in such forms that the variational geometric flow equations derived from (%
\ref{naffunctpfh}) can be written as a $\tau $-family of R-flux deformed
Einstein equations, 
\begin{equation}
\ ^{\shortmid }\widehat{\mathbf{R}}_{\ \ \gamma _{s}}^{\beta _{s}}(\tau )={%
\delta }_{\ \ \gamma _{s}}^{\beta _{s}}\ _{s}^{\shortmid }\widehat{\Im }%
^{\star }(\tau ),  \label{nonassocrffh}
\end{equation}%
where the data $\ _{s}^{\shortmid }\widehat{\Im }^{\star }(\tau )$ are
considered as some families of generating sources. For nonlinear equations (%
\ref{nonassocrffh}), the nonholonomic structure is chosen for some frame
transforms $\ ^{\shortmid }\widehat{\Im }_{\alpha ^{\prime }\beta ^{\prime
}}^{\star }=e_{\ \alpha ^{\prime }}^{\alpha _{s}}e_{\ \beta ^{\prime
}}^{\beta _{s}}\ ^{\shortmid }\widehat{\Im }_{\alpha _{s}\beta _{s}}^{\star
} $ with re-defined effective sources including distortions when $\
^{\shortmid }\widehat{\Im }_{\alpha \beta }^{\star }(\tau )=-\ ^{\shortmid }%
\widehat{\mathbf{K}}_{\alpha \beta }(\tau )-\frac{1}{2}\partial _{\tau }\
^{\shortmid }\mathbf{g}_{\alpha \beta }(\tau ))$ is parameterized in the
form 
\begin{equation}
\ ^{\shortmid }\widehat{\Im }_{\star \ \beta _{s}}^{\alpha _{s}}~(\tau ,\
^{\shortmid }u^{\gamma _{s}})=[~_{1}^{\shortmid }\widehat{\Im }^{\star
}(\kappa ,\tau ,x^{k_{1}})\delta _{i_{1}}^{j_{1}},~_{2}^{\shortmid }\widehat{%
\Im }^{\star }(\kappa ,\tau ,x^{k_{1}},y^{c_{2}})\delta
_{b_{2}}^{a_{2}},~_{3}^{\shortmid }\widehat{\Im }^{\star }(\kappa ,\tau
,x^{k_{2}},p_{c_{3}})\delta _{a_{3}}^{b_{3}},~_{4}^{\shortmid }\widehat{\Im }%
^{\star }(\kappa ,\tau ,~^{\shortmid }x^{k_{3}},p_{c_{4}})\delta
_{a_{4}}^{b_{4}}],  \label{cannonsymparamc2b}
\end{equation}%
i.e. $\ ^{\shortmid }\widehat{\Im }_{_{\beta _{s}\gamma _{s}}}^{\star }(\tau
)=diag\{\ _{s}^{\shortmid }\widehat{\Im }^{\star }(\tau )\}.$ Prescribing
certain effective sources $\ _{s}^{\shortmid }\widehat{\Im }^{\star }(\tau
)=\ _{s}^{\shortmid }\widehat{\Im }^{\star }~(\tau ,\ ^{\shortmid }u^{\gamma
_{s}}),$ we impose s-shell nonholonomic constraints for $\tau $-derivatives
of the metrics s-coefficients $\partial _{\tau }\ ^{\shortmid }\mathbf{g}%
_{\alpha _{s}\beta _{s}}(\tau )$ (which are typical for geometric flow
models) and distortions $\ ^{\shortmid }\widehat{\mathbf{K}}_{\alpha \beta
}(\tau ).$ In a similar form, we can include in $\ ^{\shortmid }\widehat{\Im 
}_{\alpha _{s}\beta _{s}}^{\star }$ nontrivial contributions by other
(effective) matter fields. The generating sources $\ _{s}^{\shortmid }%
\widehat{\Im }^{\star }(\tau )$ (\ref{cannonsymparamc2b}) can be considered
as certain symbols for parametric sources which can be re-defined in
explicit form for recurrent computations after a class of off-diagonal
solutions was found in an explicit form. For a fixed $\tau _{0}$ , the
system (\ref{nonassocrffh}) transforms into of nonlinear PDEs defining
nonholonomic Ricci solitons with self-similar configurations \cite%
{partner02,partner04,partner05,partner06}.

\section{Off-diagonal parametric solutions of nonassociative geometric flows
and gravitational field equations}

\label{sec03}In \cite{partner02,partner04,partner05,partner06}, we proved
that the system of nonlinear PDEs (\ref{nonassocrffh}) can be decoupled and
integrated in certain off-diagonal forms using the AFCDM. A number of
physically important solutions encoding nonassociative data were constructed
and discussed. Corresponding classes of solutions are generic off-diagonal
and may depend, in principle, on all spacetime and phase space local
coordinates. Technically, it is more simple to generate solutions with (at
least) one spacetime Killing symmetry. In this work, we provide formulas for
quasi-stationary solutions with Killing symmetry on $\ ^{\shortmid }\mathbf{e%
}_{4}=\partial _{4}$ when the geometric s-objects on shells $s=1,2$ do not
depend on the time-like coordinate $\ ^{\shortmid}u^{4}=y^{4}=t.$ In
abstract geometric form, the quasi-stationary solutions can be transformed
into locally anisotropic cosmological ones, with Killing symmetry on $\
^{\shortmid }\mathbf{e}_{3}=\partial _{3}$ using duality properties and
transforms of formulas for ($\ ^{\shortmid }\mathbf{e}_{4}\leftrightarrow \
^{\shortmid }\mathbf{e}_{3}$ etc.).

\subsection{Quasi-stationary off-diagonal parametric solutions}

The ansatz for generating quasi-stationary solutions of nonassociative
geometric flow equations can be chosen as {\small 
\begin{eqnarray}
d\widehat{s}^{2}(\tau ) &=&g_{i_{1}}(\tau
,x^{k_{1}})(dx^{i_{1}})^{2}+g_{a_{2}}(\tau ,x^{i_{1}},y^{3})(\mathbf{e}%
^{a_{2}}(\tau ))^{2}+\ ^{\shortmid }g^{a_{3}}(\tau ,x^{i_{2}},p_{6})(\
^{\shortmid }\mathbf{e}_{a_{3}}(\tau ))^{2}+\ ^{\shortmid }g^{a_{4}}(\tau ,\
^{\shortmid }x^{i_{3}},p_{7})(\ ^{\shortmid }\mathbf{e}_{a_{4}}(\tau ))^{2},%
\mbox{where }  \notag \\
\mathbf{e}^{a_{2}}(\tau ) &=&dy^{a_{2}}+N_{k_{1}}^{a_{2}}(\tau
,x^{i_{1}},y^{3})dx^{k_{1}},\ ^{\shortmid }\mathbf{e}_{a_{3}}(\tau
)=dp_{a_{3}}+\ ^{\shortmid }N_{a_{3}k_{2}}(\tau ,x^{i_{2}},p_{5})dx^{k_{2}},
\label{ans1rf} \\
\ ^{\shortmid }\mathbf{e}_{a_{4}}(\tau ) &=&dp_{a_{4}}+\ ^{\shortmid
}N_{a_{4}k_{3}}(\tau ,\ ^{\shortmid }x^{i_{3}},p_{7})d\ ^{\shortmid
}x^{k_{3}},\mbox{ for }
k_{1}=1,2;k_{2}=1,2,...4;k_{3}=1,2,...6;a_{2}=3,4;a_{3}=5,6;a_{4}=7,8.\  
\notag
\end{eqnarray}%
} We generate quasi-statonary solutions of (\ref{nonassocrffh}) for
nonassociative $\kappa $-parametric geometric flows for such general
parameterizations of coefficients (\ref{ans1rf}) (we can consider necessary
smooth class and even singular functions):

\begin{equation}
g_{1}(\tau ) =g_{2}(\tau )=e^{\psi (\hbar ,\kappa ;\tau
,x^{k_{1}})};g_{3}(\tau )=\frac{[\partial _{3}(\ _{2}\Psi (\tau ))]^{2}}{4(\
_{2}^{\shortmid }\widehat{\Im }^{\star }(\tau ))^{2}\{g_{4}^{[0]}(\tau
)-\int dy^{3}\frac{\partial _{3}[(\ _{2}\Psi (\tau ))^{2}]}{4(\
~_{2}^{\shortmid }\widehat{\Im }^{\star }(\tau ))}\}};  \label{qstcoef}
\end{equation}

\begin{equation}
g_{4}(\tau ) =g_{4}^{[0]}(\tau )-\int dy^{3}\frac{\partial _{3}[(\ _{2}\Psi
(\tau ))^{2}]}{4(~_{2}^{\shortmid }\widehat{\Im }^{\star }(\tau ))};\
^{\shortmid }g^{5}(\tau )=\frac{[\ ^{\shortmid }\partial ^{5}(\
_{3}^{\shortmid }\Psi (\tau ))]^{2}}{4(~_{3}^{\shortmid }\widehat{\Im }%
^{\star })^{2}\{\ ^{\shortmid }g_{[0]}^{6}(\tau )-\int dp_{5}\frac{\
^{\shortmid }\partial ^{6}[(\ _{3}^{\shortmid }\Psi (\tau ))^{2}]}{%
4(~_{3}^{\shortmid }\widehat{\Im }^{\star }(\tau ))}\}};  \notag
\end{equation}
\begin{eqnarray*}
\ ^{\shortmid }g^{6}(\tau ) &=&\ ^{\shortmid }g_{[0]}^{6}(\tau )-\int dp_{5}%
\frac{\ ^{\shortmid }\partial ^{5}[(\ _{3}^{\shortmid }\Psi (\tau ))^{2}]}{%
4(\ _{3}^{\shortmid }\widehat{\Im }^{\star })};\ ^{\shortmid }g^{7}(\tau )=%
\frac{[\ ^{\shortmid }\partial ^{7}(\ _{4}^{\shortmid }\Psi (\tau ))]^{2}}{%
4(~_{4}^{\shortmid }\widehat{\Im }^{\star }(\tau ))^{2}\{\ ^{\shortmid
}g_{[0]}^{8}(\tau )-\int dp_{7}\frac{\ ^{\shortmid }\partial ^{7}[(\
_{4}^{\shortmid }\Psi (\tau ))^{2}]}{4(~_{4}^{\shortmid }\widehat{\Im }%
^{\star }(\tau ))}\}};  \notag \\
\ ^{\shortmid }g^{8}(\tau ) &=&^{\shortmid }g_{[0]}^{8}(\tau )-\int dp_{7}%
\frac{\ ^{\shortmid }\partial ^{7}[(\ _{4}^{\shortmid }\Psi (\tau ))^{2}]}{%
4(\ _{4}^{\shortmid }\widehat{\Im }^{\star }(\tau ))},%
\mbox{ and for
nontrivial N-connection coefficients: }  \notag
\end{eqnarray*}
\begin{eqnarray*}
N_{k_{1}}^{3}(\tau ) &=&\frac{\partial _{k_{1}}(\ _{2}\Psi )}{\partial
_{3}(\ _{2}\Psi )};N_{k_{1}}^{4}(\tau )=\ _{1}n_{k_{1}}(\tau )+\
_{2}n_{k_{1}}(\tau )\int dy^{3}\frac{\partial _{3}[(\ _{2}\Psi )^{2}]}{4(\
~_{2}^{\shortmid }\widehat{\Im }^{\star })^{2}|g_{4}^{[0]}(\tau )-\int dy^{3}%
\frac{\partial _{3}[(\ _{2}\Psi )^{2}]}{4(~_{2}^{\shortmid }\widehat{\Im }%
^{\star })}|^{5/2}}; \\
\ ^{\shortmid }N_{5k_{2}}(\tau ) &=&\frac{\partial _{k_{2}}(\
_{3}^{\shortmid }\Psi )}{\ ^{\shortmid }\partial ^{5}(\ _{3}^{\shortmid
}\Psi )};\ ^{\shortmid }N_{6k_{2}}(\tau )=\ _{1}^{\shortmid }n_{k_{2}}(\tau
)+\ _{2}^{\shortmid }n_{k_{2}}(\tau )\int dp_{5}\frac{\ ^{\shortmid
}\partial ^{5}[(\ _{3}^{\shortmid }\Psi )^{2}]}{4(\ _{3}^{\shortmid }%
\widehat{\Im }^{\star })^{2}|\ ^{\shortmid }g_{[0]}^{6}(\tau )-\int dp_{5}%
\frac{\ ^{\shortmid }\partial ^{5}[(\ _{3}^{\shortmid }\Psi )^{2}]}{4(\
_{3}^{\shortmid }\widehat{\Im }^{\star })}|^{5/2}}; \\
\ ^{\shortmid }N_{7k_{3}}(\tau ) &=&\frac{\partial _{k_{3}}(\
_{4}^{\shortmid }\Psi )}{\ ^{\shortmid }\partial ^{7}(\ _{4}^{\shortmid
}\Psi )};\ ^{\shortmid }N_{8k_{3}}(\tau )=\ _{1}^{\shortmid }n_{k_{3}}(\tau
)+\ _{2}^{\shortmid }n_{k_{3}}(\tau )\int dp_{7}\frac{\ ^{\shortmid
}\partial ^{7}[(\ _{4}^{\shortmid }\Psi )^{2}]}{4(\ _{4}^{\shortmid }%
\widehat{\Im }^{\star })^{2}|\ ^{\shortmid }g_{[0]}^{8}(\tau )-\int dp_{7}%
\frac{\ ^{\shortmid }\partial ^{7}[(\ _{4}^{\shortmid }\Psi )^{2}]}{4(\
_{4}^{\shortmid }\widehat{\Im }^{\star })}|^{5/2}}.
\end{eqnarray*}%
The s-adapted coefficients (\ref{qstcoef}) are determined in general form by
respective generating and integration functions with $\tau $-parametric
dependence: 
\begin{eqnarray}
\mbox{generating functions: } &&\psi (\tau )\simeq \psi (\hbar ,\kappa ;\tau
,x^{k_{1}});\ _{2}\Psi (\tau )\simeq \ _{2}\Psi (\hbar ,\kappa ;\tau
,x^{k_{1}},y^{3});  \label{integrfunctrf} \\
&&\ _{3}^{\shortmid }\Psi (\tau )\simeq \ _{3}^{\shortmid }\Psi (\hbar
,\kappa ;\tau ,x^{k_{2}},p_{5});\ _{4}^{\shortmid }\Psi (\tau )\simeq \
_{4}^{\shortmid }\Psi (\hbar ,\kappa ;\tau ,\ ^{\shortmid }x^{k_{3}},p_{7});
\notag \\
\mbox{generating sources: \quad} &&\ _{1}^{\shortmid }\widehat{\Im }^{\star }%
\mathcal{(\tau )}\simeq \ ~_{1}^{\shortmid }\widehat{\Im }^{\star }(\hbar
,\kappa ;\tau ,x^{k_{1}});\ ~_{2}^{\shortmid }\widehat{\Im }^{\star }%
\mathcal{(\tau )}\simeq \ ~_{2}^{\shortmid }\widehat{\Im }^{\star }(\hbar
,\kappa ;\tau ,x^{k_{1}},y^{3});  \notag \\
&&\ _{3}^{\shortmid }\widehat{\Im }^{\star }\mathcal{(\tau )}\simeq \
~_{3}^{\shortmid }\widehat{\Im }^{\star }(\hbar ,\kappa ;\tau
,x^{k_{2}},p_{5});~_{4}^{\shortmid }\widehat{\Im }^{\star }\mathcal{(\tau )}%
\simeq \ ~~_{4}^{\shortmid }\widehat{\Im }^{\star }(\hbar ,\kappa ;\tau ,\
^{\shortmid }x^{k_{3}},p_{7});  \notag \\
\mbox{integrating functions: } &&  \notag \\
g_{4}^{[0]}(\tau ) &\simeq &g_{4}^{[0]}(\hbar ,\kappa ;\tau ,x^{k_{1}}),\
_{1}n_{k_{1}}\mathcal{(\tau )}\simeq \ _{1}n_{k_{1}}(\hbar ,\kappa ;\tau
,x^{j_{1}}),\ _{2}n_{k_{1}}\mathcal{(\tau )}\simeq \ _{2}n_{k_{1}}(\hbar
,\kappa ;\tau ,x^{j_{1}});  \notag \\
\ ^{\shortmid }g_{[0]}^{6}\tau ) &\simeq &\ ^{\shortmid }g_{[0]}^{6}(\hbar
,\kappa ;\tau ,x^{k_{2}}),\ _{1}n_{k_{2}}\mathcal{(\tau )}\simeq \
_{1}n_{k_{2}}(\hbar ,\kappa ;\tau ,x^{j_{2}}),\ _{2}n_{k_{2}}\mathcal{(\tau )%
}\simeq \ _{2}n_{k_{2}}(\hbar ,\kappa ;\tau ,x^{j_{2}});  \notag \\
\ ^{\shortmid }g_{[0]}^{8}(\tau ) &\simeq &\ ^{\shortmid }g_{[0]}^{8}(\hbar
,\kappa ;\tau ,\ ^{\shortmid }x^{j_{3}}),\ _{1}n_{k_{3}}\mathcal{(\tau )}%
\simeq \ _{1}^{\shortmid }n_{k_{3}}(\hbar ,\kappa ;\tau ,\ ^{\shortmid
}x^{j_{3}}),\ _{2}n_{k_{3}}\mathcal{(\tau )}\simeq \ _{2}^{\shortmid
}n_{k_{3}}(\hbar ,\kappa ;\tau ,\ ^{\shortmid }x^{j_{3}});  \notag
\end{eqnarray}%
and $\psi (\tau )$ are solutions of a respective family of 2-d Poisson
equations, $\partial _{11}^{2}\psi (\hbar ,\kappa ;\tau ,x^{k_{1}})+\partial
_{22}^{2}\psi (\hbar ,\kappa ;\tau ,x^{k_{1}})=2\ _{1}\widehat{\Im }^{\star
}(\hbar ,\kappa ;\tau ,x^{k_{1}}),$ encoding geometric flows of
nonassociative data if, in general, $\ _{1}\widehat{\Im }^{\star }\mathcal{%
(\tau )}$ contains such nonholonomic parametric dependencies.

\subsection{Nonlinear symmetries and $\protect\tau $-running cosmological
constants}

Quasi-stationary s-metric (\ref{ans1rf}) with N-connection coefficients (\ref%
{qstcoef}) posses very important nonlinear symmetries, see details in \cite%
{partner02,partner04,partner05,partner06}. Such nonlinear transforms allow
to construct nonassociative nonholonomic geometric flow deformations of
families of \textbf{prime} s-metrics $\ _{s}^{\shortmid }\mathbf{\mathring{g}%
}(\tau )$ into a corresponding family of \textbf{target} s-metrics $\
_{s}^{\shortmid }\mathbf{g}(\tau )$ defining a nonassociative geometric flow
evolution scenarios of quasi-stationary metrics on $\ _{s}^{\star }\mathcal{M%
},$ 
\begin{equation}
\ _{s}^{\shortmid }\mathbf{\mathring{g}}(\tau )\rightarrow \ _{s}^{\shortmid
}\mathbf{g}(\tau )=[\ ^{\shortmid }g_{\alpha _{s}}(\tau )=\ ^{\shortmid
}\eta _{\alpha _{s}}(\tau )\ ^{\shortmid }\mathring{g}_{\alpha _{s}}(\tau
),\ ^{\shortmid }N_{i_{s-1}}^{a_{s}}(\tau )=\ ^{\shortmid }\eta
_{i_{s-1}}^{a_{s}}(\tau )\ ^{\shortmid }\mathring{N}_{i_{s-1}}^{a_{s}}(\tau
)].  \label{offdiagdefr}
\end{equation}%
The gravitational $\eta $-polarization functions in (\ref{offdiagdefr}) have
to be defined in a form that $\ _{s}^{\shortmid }\mathbf{g}(\tau )$ define
certain classes of exact/parametric solutions when $\ _{s}^{\shortmid }%
\mathbf{\mathring{g}}(\tau )$ can be arbitrary ones, or chosen as cetain
physically important solutions of some (modified) Einstein equations in a
MGT or GR. Using such nonlinear symmetries, we can re-define the geometric
flows of generating functions, generating sources and integration functions (%
\ref{integrfunctrf}) and relate them to certain the effective sources shell $%
\tau $-running cosmological constants $\ _{s}^{\shortmid }\Lambda (\tau ).$
Conventionally, we parameterize such nonlinear transforms as 
\begin{eqnarray}
&&(\ _{s}\Psi (\tau ),\ _{s}^{\shortmid }\widehat{\Im }^{\star }(\tau
))\leftrightarrow (\ _{s}^{\shortmid }\mathbf{g}(\tau ),\ _{s}^{\shortmid }%
\widehat{\Im }^{\star }(\tau ))\leftrightarrow (\ _{s}^{\shortmid }\eta
(\tau )\ ^{\shortmid }\mathring{g}_{\alpha _{s}}(\tau )\sim (\ ^{\shortmid
}\zeta _{\alpha _{s}}(\tau )(1+\kappa \ ^{\shortmid }\chi _{\alpha
_{s}}(\tau ))\ ^{\shortmid }\mathring{g}_{\alpha _{s}}(\tau ),\
_{s}^{\shortmid }\widehat{\Im }^{\star }(\tau ))\leftrightarrow  \notag \\
&&(\ _{s}\Phi (\tau ),\ _{s}^{\shortmid }\Lambda (\tau ))\leftrightarrow (\
_{s}^{\shortmid }\mathbf{g},\ _{s}^{\shortmid }\Lambda (\tau
))\leftrightarrow (\ _{s}^{\shortmid }\eta (\tau )\ ^{\shortmid }\mathring{g}%
_{\alpha _{s}}(\tau )\sim (\ ^{\shortmid }\zeta _{\alpha _{s}}(\tau
)(1+\kappa \ ^{\shortmid }\chi _{\alpha _{s}}(\tau ))\ ^{\shortmid }%
\mathring{g}_{\alpha _{s}}(\tau ),\ _{s}^{\shortmid }\Lambda (\tau )),
\label{nonlinsymr}
\end{eqnarray}%
where $\ _{s}^{\shortmid }\Lambda _{0}=\ _{s}^{\shortmid }\Lambda (\tau
_{0}) $ can be fixed for nonassociative Ricci soliton symmetries.

In explicit form, the nonlinear transforms (\ref{nonlinsymr}) are descibed
by s-adapted formulas, when 
\begin{eqnarray}
\partial _{3}[(\ _{2}\Psi )^{2}] &=&-\int dy^{3}~_{2}^{\shortmid }\widehat{%
\Im }^{\star }\partial _{3}g_{4}\simeq -\int dy^{3}~_{2}^{\shortmid }%
\widehat{\Im }^{\star }\partial _{3}(\ ^{\shortmid }\eta _{4}\ \mathring{g}%
_{4})\simeq -\int dy^{3}~_{2}^{\shortmid }\widehat{\Im }^{\star }\partial
_{3}[\ ^{\shortmid }\zeta _{4}(\tau )(1+\kappa \ ^{\shortmid }\chi _{4}(\tau
))\ \mathring{g}_{4}],  \notag \\
(\ _{2}\Phi )^{2} &=&-4\ _{2}\Lambda (\tau )g_{4}\simeq -4\ _{2}\Lambda
(\tau )\ ^{\shortmid }\eta _{4}\ \mathring{g}_{4}\simeq -4\ _{2}\Lambda
(\tau )\ ^{\shortmid }\zeta _{4}(\tau )(1+\kappa \ ^{\shortmid }\chi
_{4}(\tau ))\ \mathring{g}_{4};  \notag
\end{eqnarray}%
\begin{eqnarray*}
~\ ^{\shortmid }\partial ^{5}[(\ _{3}^{\shortmid }\Psi )^{2}] &=&-\int
dp_{5}~_{3}^{\shortmid }\widehat{\Im }^{\star }\ ^{\shortmid }\partial ^{5}\
^{\shortmid }g^{6}\simeq -\int dp_{5}~_{3}^{\shortmid }\widehat{\Im }^{\star
}\ ^{\shortmid }\partial ^{5}(\ ^{\shortmid }\eta ^{6}\ ^{\shortmid }%
\mathring{g}^{6})\simeq -\int dp_{5}~_{3}^{\shortmid }\widehat{\Im }^{\star
}\ ^{\shortmid }\partial ^{5}[\ ^{\shortmid }\zeta ^{6}(\tau )(1+\kappa \
^{\shortmid }\chi ^{6}(\tau ))\ \mathring{g}^{6}], \\
(\ _{3}^{\shortmid }\Phi )^{2} &=&-4\ _{3}^{\shortmid }\Lambda (\tau )\
^{\shortmid }g^{6}\simeq \ -4\ _{3}^{\shortmid }\Lambda (\tau )\ ^{\shortmid
}\eta ^{6}(\tau )\ ^{\shortmid }\mathring{g}^{6}\simeq -4\ _{3}^{\shortmid
}\Lambda (\tau )\ ^{\shortmid }\zeta ^{6}(\tau )(1+\kappa \ ^{\shortmid
}\chi ^{6}(\tau ))\ ^{\shortmid }\mathring{g}^{6};
\end{eqnarray*}%
\begin{eqnarray*}
~\ ^{\shortmid }\partial ^{7}[(\ _{4}^{\shortmid }\Psi )^{2}] &=&-\int
dp_{7}~_{4}^{\shortmid }\widehat{\Im }^{\star }\ ^{\shortmid }\partial ^{7}\
^{\shortmid }g^{8}\simeq -\int dp_{7}~_{4}^{\shortmid }\widehat{\Im }^{\star
}\ ^{\shortmid }\partial ^{7}(\ ^{\shortmid }\eta ^{8}\ ^{\shortmid }%
\mathring{g}^{8})\simeq -\int dp_{7}~_{4}^{\shortmid }\widehat{\Im }^{\star
}\ ^{\shortmid }\partial ^{7}[\ ^{\shortmid }\zeta ^{8}(\tau )(1+\kappa \
^{\shortmid }\chi ^{8}(\tau ))\ \mathring{g}^{8}], \\
(\ _{4}^{\shortmid }\Phi )^{2} &=&-4\ _{4}^{\shortmid }\Lambda (\tau )\
^{\shortmid }g^{8}\simeq \ -4\ _{4}^{\shortmid }\Lambda (\tau )\ ^{\shortmid
}\eta ^{8}(\tau )\ ^{\shortmid }\mathring{g}^{8}\simeq -4\ _{4}^{\shortmid
}\Lambda (\tau )\ ^{\shortmid }\zeta ^{8}(\tau )(1+\kappa \ ^{\shortmid
}\chi ^{8}(\tau ))\ ^{\shortmid }\mathring{g}^{8}.
\end{eqnarray*}

The nonlinear symmetries (\ref{nonlinsymr}) when $\ [\ _{s}\Psi (\tau ),\
_{s}^{\shortmid }\widehat{\Im }^{\star }(\tau )]\rightarrow \lbrack \
_{s}\Phi (\tau ),\ _{s}^{\shortmid }\Lambda (\tau )]$ allow us to change the
generating data and construct different types of off-diagonal solutions
which encoded effective and matter field sources in explicit form, or such
(in general, nonassociative contributions) are encoded into off-diagonal
terms with effective $\tau $-running cosmological constants.

The $\kappa $-linear parametric nonassociative geometric flow equations (\ref%
{nonassocrffh}) (which can be written a integrated in quasi-stationary form
using $\ _{s}\Psi (\tau )$) can be re-defined equivalently as a system of
functional equations with $\ _{s}\Phi (\tau )$ for certain shell effective $%
\tau $-running constants $\ _{s}^{\shortmid }\Lambda (\tau )$ introduced for
modelling geometric flow evolution processes. In functional form, the
nonassociative geometric flow equations can be written in cetain equivalent
forms as $\ ^{\shortmid }\widehat{\mathbf{R}}_{\ \ \gamma _{s}}^{\beta
_{s}}(\tau ,\ _{s}\Phi (\tau ),\ _{s}^{\shortmid }\widehat{\Im }^{\star
}(\tau ))={\delta }_{\ \ \gamma _{s}}^{\beta _{s}}\ _{s}^{\shortmid }\Lambda
(\tau ).$ Usually, it is more easy technically to generate solutions of such
systems of nonlinear PDEs and their physical interpretation can be related
to certain classes of of solutions with effective cosmological constants.
Any deformations $(\ _{s}^{\shortmid }\eta (\tau )\ ^{\shortmid }\mathring{g}%
_{\alpha _{s}}(\tau ))$ allow to construct families of non-perturbative
deformations of some prime metrics when $\ ^{\shortmid }\mathring{g}_{\alpha
_{s}}(\tau )$ are additionally defined by some physically important
constants (for instance, the BH mass and charge), which may be with small
polarizations in term of $(\ ^{\shortmid }\zeta _{\alpha _{s}}(\tau
)(1+\kappa \ ^{\shortmid }\chi _{\alpha _{s}}(\tau ))\ ^{\shortmid }%
\mathring{g}_{\alpha _{s}}(\tau ),\ _{s}^{\shortmid }\Lambda (\tau )),$ or
self-consistently imbedded into an a nonlinear phase space vacuum. Another
class of nonassociative geometric flow and off-diagonal deformations may
result into deformations of certain horizon deformations (for instance,
transforming spherical/toroid data into certain ellipsoid and
ellpsoid-toroid configurations) as we discused in \cite%
{partner04,partner05,partner06}.

\section{Functional renormalization of nonassociative geometric flows \&
gravity}

\label{sec04}In this section we show how Weinberg's asymptotic safety
scenario \cite{weinberg79,niedermaier06,gies16} can be generalized for
constructing a QG theory with embedding nonassociative R-flux data from
string theory. In explicit form, our approach is based on parametric
decompositions and off-diagonal solutions of nonassociative geometric flow
and Ricci soliton (generalized Einstein) equations constructed in explicit
form as in previous section and characterized by respective nonlinear
symmetries (\ref{nonlinsymr}). We elaborate a new nonholonomic functional
renormalization group techniques to determine the RG flows on phase spaces
modelled as cotangent Lorentz bundle when the total phase space two-loop
counter term is defined as a canonical generalization of the GR term found
by Goroff and Sagnotti \cite{goroff1,goroff2,ven92}.

\subsection{Nonholonomic phase space functional renormalization}

Powerful tools to investigate renormalizability based on interacting RG
fixed points using functional RG were elaborated with generalizations to
off-diagonal heat-kernel methods in \cite{wett93,dec06,ans07,gies16}. The
approaches were based on finding certain interacting RG fixed points
provided by the functional RG and respective RG equation, 
\begin{equation}
\check{p}\partial _{\check{p}}\Gamma _{\check{p}}(\tau )=\frac{1}{2}%
Str[(\Gamma _{\check{p}}^{(2)}(\tau )+\Re _{\check{p}}(\tau ))^{-1}\check{p}%
]\partial _{\check{p}}\Re _{\check{p}}(\tau )], \mbox{ on } \ ^{\shortmid }%
\mathcal{M}^{\star }.  \label{frgreq}
\end{equation}%
The $\tau $-family of parametric functional equation (\ref{frgreq}) realizes
Wilson's idea of renormalization by integrating out quantum fluctuations in
a nonholonomic dyadic shell-by-shell manner as we considered in sections \ref%
{sec02} and \ref{sec03}.\footnote{%
We use a different system of notations with $\check{p}$ instead of $k$
considered, for instance in \cite{gies16} because a similar symbol $\kappa =%
\mathit{\ell }_{s}^{3}/6\hbar $ is used for defining the twisted star
product (\ref{starpn}).} The two-point correlators $\Gamma _{\check{p}%
}^{(2)}(\tau )$ will be defined in a form to result in a formally exact
equation when the regulators $\Re _{\check{p}}(\tau )$ will describe changes
driven by phase space quantum fluctuations with momenta $p_{a}$ close to a $%
\check{p}=\{\check{p}_{a}\}$ and their evolution as nonassociative geometric
flows. Such values will be defined in terms of canonical (hat) geometric
variables encoding $k$-parametric nonassociative R-flux data as certain
off-diagonal solutions of (\ref{nonassocrffh}). Those solutions are
nonholonomically adapted to result in solutions of nonassocive parametric RG
equations (\ref{frgreq}) encoding also nonlinear symmetries (\ref{nonlinsymr}%
).

Three key advantages of the nonassociative RG equations (\ref{frgreq}) are
that:\ 1) we can begin with approximations on some small parameters $\hbar $
and $\kappa $ (and other parameters, for instance, defining BH or WH
solutions, or certain cosmological scenarios); but both the 2) RG flows and
3) nonassociative geometric flow scenarios can be elaborated without small
expansion parameters. This means, for instance, that we can work with
general gravitational $\eta $-polarization functions in (\ref{offdiagdefr})
in a form adapted to nonlinear symmetries (\ref{nonlinsymr}) preserving the
chosen type of geometric and quantum evolution scenarios for a class of
off-diagonal solutions. More than that: 4) we do not need to specify a
fundamental action ap priori which is very important because it is not
possible to elaborate a general nonassociative variational calculus (in
principle, we can define an infinite number of nonassociative or
noncommutative differential/ integral calculi), see details in \cite%
{partner05,partner06}. The F- and W-functionals (\ref{naffunctpfh}) can be
computed in parametric form for any class of off-diagonal solutions. Fixing $%
\tau =\tau _{0},$ we can derive variational phase-modified Einstein
equations which are equivalent to certain Ricci soliton equations. All
mentioned properties 1-4) allow us to formulate a procedure predestined for
searching for, or fixing certain type of fixed points of the renormalization
flow and nonassociative geometric evolution flow beyond the realm of
perturbation theory and for a various class of thermo-field and geometric
thermo-evolution models.

For simplicity, the rest part of the paper focus on the case when the
nonassociative geometric evolution is carried out by a (full) phase space
s-metric $\ ^{\shortmid }\mathbf{g}_{\alpha _{s}\beta _{s}}^{\star }= $ $\
^{\shortmid }\underline{\mathbf{g}}_{\alpha _{s}\beta _{s}}^{\star }+\
^{\shortmid }\mathbf{h}_{\alpha _{s}\beta _{s}}^{\star },$ where the
background s-metric $\ ^{\shortmid }\underline{\mathbf{g}}_{\alpha _{s}\beta
_{s}}^{\star }$ and fluctuations $\ ^{\shortmid }\mathbf{h}_{\alpha
_{s}\beta _{s}}^{\star }$ can be considered in some forms when both $\
^{\shortmid }\mathbf{g}_{\alpha _{s}\beta _{s}}^{\star }$ and $\ ^{\shortmid}%
\underline{\mathbf{g}}_{\alpha _{s}\beta _{s}}^{\star }$ are solutions of (%
\ref{nonassocrffh}). We study a nonassociative geometric flow RG models
generalizing the gravitational renormalization flow for the Einstein-Hilbert
action being R-flux deformed on the phase spaces in canonical s-adapted
variables (following \cite{ven92} when the formulas are modified to hat
geometric objects). For (\ref{frgreq}), we write 
\begin{equation}
\Gamma _{\check{p}}(\tau )=\ \ _{\check{p}}^{\shortmid }\widehat{\mathcal{W}}%
_{\kappa }^{\star }(\tau )+\ \ _{\check{p}}^{\shortmid }\widehat{\Gamma }%
_{\kappa }^{GS}(\tau ),\mbox{ for }\ \ _{\check{p}}^{\shortmid }\widehat{%
\Gamma }_{\kappa }^{GS}(\tau )=\underline{\sigma }_{\check{p}%
}\int_{^{\shortmid }\Xi }\ ^{\shortmid }\widehat{\mathbf{C}}_{\alpha \beta
}^{\star \quad \mu \nu }(\tau )\ ^{\shortmid }\widehat{\mathbf{C}}_{\mu \nu
}^{\star \quad \rho \sigma }(\tau )\ ^{\shortmid }\widehat{\mathbf{C}}_{\rho
\sigma }^{\star \quad \alpha \beta }(\tau )e^{-\ ^{\shortmid }\widehat{f}%
(\tau )}\ d\ ^{\shortmid }\mathcal{V}ol(\tau ),  \label{effectact}
\end{equation}%
where $\ _{\check{p}}^{\shortmid }\widehat{\mathcal{W}}_{\kappa
}^{\star}(\tau )$ is the W-functional (\ref{naffunctpfh}) in a 8-d region $%
^{\shortmid }\Xi \subset \ ^{\shortmid }\mathcal{M}^{\star }$ close to a $%
\check{p}$ and $\ ^{\shortmid }\widehat{\mathbf{C}}_{\alpha \beta }^{\star
\quad \mu \nu }$ being the Weyl d-tensor defined by the d-connection $\
^{\shortmid }\widehat{\mathbf{D}}^{\star }.$ The scale-dependent phase space
coupling $\underline{\sigma }_{\check{p}}$ is chosen in a form that for
spacetime projections diverges at least as $\ln \check{p}$ for $\check{p}%
\rightarrow \infty $ even in flat-space on-shell limit $\Lambda _{\check{p}%
}\rightarrow 0$ and after the Newton coupling has been renormalized. For a
fixed $\tau _{0},$ we consider that the $s=1,2$ part of the effective
average action is supplemented by a standard gauge-fixing procedure and
harmonic gauge used in \cite{reuter98}.

The nonholonomic canonical RG flow equations for couplings (in general, we
can consider various parameters and constants defining off-diagonal
solutions of type (\ref{ans1rf}) which may define quasi-stationary, or in
dual form, locally anisotropic phase space configurations) can be found by
substituting the functionals (\ref{effectact}) into (\ref{frgreq}) and
computing the respective d- or s-adapted coefficients of curvature terms.
The evaluation of the trace utilized the same universal RG technics together
with off-diagonal heat-kernel methods \cite{wett93,dec06,ans07,gies16} has
to be combined with the nonassociative AFCDM \cite%
{partner02,partner04,partner05,partner06}.

Using nonholonomic geometric data $(\ ^{\shortmid }\mathbf{g}_{\alpha
_{s}\beta _{s}}^{\star },\ ^{\shortmid }\widehat{\mathbf{D}}_{\gamma
_{s}}^{\star })$ and additional frame transforms $\ ^{\shortmid }\mathbf{g}%
_{\alpha ^{\prime }\beta ^{\prime }}^{\star }=e_{\ \alpha ^{\prime
}}^{\alpha _{s}}e_{\ \beta ^{\prime }}^{\beta _{s}}\ ^{\shortmid }\widehat{%
\mathbf{g}}_{\alpha _{s}\beta _{s}}^{\star },$ when a class of off-diagonal
solutions (\ref{ans1rf}) is related to a configuration of s-metrics with
zero $\ ^{\shortmid }\widehat{\mathbf{C}}_{\alpha \beta }^{\star \quad \mu
\nu }=0$ (when Lorentz spacetime projections may contain a nonzero $%
C_{ij}^{\star \quad kl}$ for a corresponding LC-connection $\ ^{\shortmid }%
\mathbf{\nabla _{k}^{\star },}$ see formulas (\ref{twoconsstar}) for
distortions of $\ ^{\shortmid }\mathbf{\nabla ^{\star }}$). This entails
that there is no feedback on the spacetime Goroff-Sagnotti term on the phase
space renormalization flow of the Newton constant, cosmological shell
constants, and other type constants defining off-diagonal phase space
nonassociative geometric flows and their solutions.\footnote{\label{fncompt}%
See similar details in \cite{gies16} and references therein on computer
methods with Mathematical package xAct, for 900 Goroff-Sagnotti terms
corresponding to one term of the Einstein-Hilbert vertex.}

\subsection{Parametric phase space beta functions, fixed points, and
nonassociative RG flows}

We consider a nonholonomic 4+4 phase space splitting when $\ ^{\shortmid }%
\widehat{\mathbf{D}}^{\star }=(h\ ^{\shortmid }\widehat{\mathbf{D}}^{\star
},\ \ c\ ^{\shortmid }\widehat{\mathbf{D}}^{\star })$ results in zero $\
^{\shortmid }\widehat{\mathcal{C}}^{\star }=\{\ ^{\shortmid }\widehat{%
\mathbf{C}}_{\ \beta \gamma \delta }^{\star \alpha }\}=0$ for nonlinear
symmetries of solutions (\ref{ans1rf}) related to $\ _{s}^{\shortmid
}\Lambda (\tau )=\{\ _{1}\Lambda (\tau )=\ _{2}\Lambda (\tau )=\Lambda (\tau
),\ _{3}^{\shortmid }\Lambda (\tau )=\ _{4}^{\shortmid }\Lambda (\tau )=\
^{\shortmid }\Lambda (\tau )\}$ and study the nonholonomic RG flows
resulting from the effective action (\ref{effectact}). Such an analysis is
convenient to be performed in terms of dimensionless couplings 
\begin{equation}
\varrho _{\breve{\imath}}(\tau )=\{\lambda (\tau ):=\ _{\check{p}}\Lambda
(\tau )\check{p}^{-2},\varrho (\tau ):=\ _{\check{p}}G(\tau )\check{p}%
^{2},\sigma (\tau ):=\underline{\sigma }_{\check{p}}\check{p}^{2}\};\
^{\shortmid }\varrho _{\breve{\imath}}(\tau )=\{\ ^{\shortmid }\lambda (\tau
):=\ _{\check{p}}^{\shortmid }\Lambda (\tau )\check{p}^{-2},\ ^{\shortmid
}\varrho (\tau ),\ ^{\shortmid }\sigma (\tau )\}.  \label{dimenssc}
\end{equation}%
In these formulas, there are considered possible $\tau $-running of the
(horizontal) cosmological constant $\ _{\check{p}}\Lambda (\tau )$ and
Newton constant $\ _{\check{p}}G(\tau )$ when similar (covertical) values $\
^{\shortmid }\lambda (\tau ),\ ^{\shortmid }\varrho (\tau ),\ ^{\shortmid
}\sigma (\tau )$ can be arbitrary ones which can be prescribed for
elaborating a realistic total phase space scenarios. Using conventions (\ref%
{dimenssc}), we express (\ref{frgreq}) in terms of respective beta functions%
\begin{equation*}
\check{p}\partial _{\check{p}}\varrho _{\breve{\imath}}(\tau )=\beta _{%
\breve{\imath}}[\lambda (\tau ),\varrho (\tau ),\sigma (\tau )]\mbox{ and }%
\check{p}\partial _{\check{p}}\ ^{\shortmid }\varrho _{\breve{\imath}}(\tau
)=\ ^{\shortmid }\beta _{\breve{\imath}}[\ ^{\shortmid }\lambda (\tau ),\
^{\shortmid }\varrho (\tau ),\ ^{\shortmid }\sigma (\tau )].
\end{equation*}%
The $\beta $-function for the dimensionless Newton's constant and
cosmological constant in GR is known from \cite{reuter98} when latter it was
introduced the so-called Litim regulator \cite{litim01} (we use and
anomalous dimension $\varsigma (\tau )$ of Newton's constant with $\tau $%
-running). For the spacetime h-components, we write respectively%
\begin{eqnarray}
\beta _{\rho }(\tau ) &=&[2+\varsigma (\tau )]\varrho (\tau )\mbox{ and }%
\beta _{\lambda }(\tau )=[\varsigma (\tau )-2]\lambda (\tau )+\frac{\varrho
(\tau )}{2\pi }(\frac{5}{1-2\lambda (\tau )}-\frac{5\varsigma (\tau )}{%
6(1-2\lambda (\tau ))}-4),\mbox{ for }  \label{betah} \\
\varsigma (\tau ) &=&\frac{\varrho (\tau )B_{1}(\tau )}{1-\varrho (\tau
)B_{2}(\tau )},\mbox{ where }B_{1}(\tau )=\frac{1}{3\pi }(\frac{5}{%
1-2\lambda (\tau )}-\frac{9}{(1-2\lambda (\tau ))^{2}}-5),B_{2}(\tau )=\frac{%
1}{6\pi (1-2\lambda (\tau ))}(\frac{3}{1-2\lambda (\tau )}-\frac{5}{2}). 
\notag
\end{eqnarray}%
The effective action (\ref{effectact}) defines also the third beta function $%
\beta _{\rho }(\tau )$ for $\sigma (\tau ),$ 
\begin{equation}
\beta _{\rho }(\tau )=c_{0}(\tau )+[2+c_{1}(\tau )]\sigma (\tau )+c_{2}(\tau
)[\sigma (\tau )]^{2}+c_{3}(\tau )[\sigma (\tau )]^{2},  \label{betasigma}
\end{equation}%
where the coefficients are computed 
\begin{eqnarray*}
c_{0}(\tau ) &=&\frac{1}{64\pi ^{2}(1-2\lambda (\tau ))}[\frac{2-\varsigma
(\tau )}{2(1-2\lambda (\tau ))}+\frac{6-\varsigma (\tau )}{(1-2\lambda (\tau
))^{3}}-\frac{5\varsigma (\tau )}{378}], \\
c_{1}(\tau ) &=&\frac{3\varrho (\tau )}{16\pi (1-2\lambda (\tau ))^{2}}%
[5(6-\varsigma (\tau ))+\frac{23(8-\varsigma (\tau ))}{8(1-2\lambda (\tau ))}%
-\frac{7(10-\varsigma (\tau ))}{10(1-2\lambda (\tau ))^{2}}], \\
c_{3}(\tau ) &=&\frac{[\varrho (\tau )]^{2}}{2(1-2\lambda (\tau ))^{3}}[%
\frac{233}{10}(12-\varsigma (\tau ))-\frac{9(14-\varsigma (\tau ))}{%
7(1-2\lambda (\tau ))}],c_{4}(\tau )=\frac{16\pi \lbrack \varrho (\tau
)]^{3}(18-\varsigma (\tau ))}{(1-2\lambda (\tau ))^{4}}.
\end{eqnarray*}%
If we fix a $\tau _{0}$ for the LC-connection, the formulas (\ref{betah})
and (\ref{betasigma}) reproduce the main results of \cite{gies16} if $%
c_{3}(\tau _{0})$ \ is positive for any admissible $\lambda (\tau _{0}),$
with an effective Newton coupling $\varrho (\tau _{0})>0,$ and $\varsigma
(\tau _{0})<18.$ We can consider the \ same results for fixed point and RG
flows which involve numeric calculations, graphs and conditions when the
Ricci soliton system exhibits on the 4-d Lorentz base manifold Gaussian a
fixed point, GEP, or a non-Gaussian, NGEP, behaviour which controls the
high-energy limit of gravity as required for the asymptotic safety scenario.
The off-diagonal and $\tau $-dependence allows to generation families of
such systems with $c_{3}(\tau )$ being gauge independent and positivity
which is independent of metric parametrization. This means that the
inclusion of the Goroff-Sagnotti term lives asymptotic safety which was
proven for the Einstein-Hilbert ansatz and R$^{2}$-extensions in a fully
intact form even for canonical distortions of the LC-connection (see
formulas (14)-(18) in \cite{gies16}).

In a more general approach for nonassociative geometric flows on phase
spaces, we can adapt the nonholonomic structure and reproduce the formulas (%
\ref{betah}) and (\ref{betasigma}) with covertical labels of type $\
^{\shortmid }\beta _{\breve{\imath}}[\ ^{\shortmid }\lambda (\tau ),\
^{\shortmid }\varrho (\tau ),\ ^{\shortmid }\sigma (\tau )]$ and respective $%
\ ^{\shortmid }\varsigma (\tau ),\ ^{\shortmid }B_{1}(\tau ),\ ^{\shortmid
}B_{2}(\tau )$ used for computing, for instance, certain 
\begin{equation*}
\ ^{\shortmid }\beta _{\rho }(\tau )=\ ^{\shortmid }c_{0}(\tau )+[2+\
^{\shortmid }c_{1}(\tau )]\ ^{\shortmid }\sigma (\tau )+\ ^{\shortmid
}c_{2}(\tau )[\ ^{\shortmid }\sigma (\tau )]^{2}+\ ^{\shortmid }c_{3}(\tau
)[\ ^{\shortmid }\sigma (\tau )]^{2}.
\end{equation*}
The cofiber asymptotic behaviour does not involve a Newton constant, but a $%
\ ^{\shortmid }\varrho (\tau )$ can be introduced effectively and controlled
by realistic nonassociative geometric evolution scenarios. Such
constructions are determined by the twisted star product encoding R-flux
deformations from string theory. The nonlinear symmetries (\ref{nonlinsymr})
point to certain nontrivial values $\ ^{\shortmid }\lambda (\tau )$ which
can be considered as NGEP determined by nonassociative parametric
contributions. So, Goroff-Sagnotti terms in the base Lorentz manifold and,
in canonical form, on the nonassociative phase space, may shift Gaussian
fixed points because of R-flux contributions and nonlinear symmetries of
nonassociative geometric off-diagonal evolution but asymptotic scenarios can
be prescribed for well-defined data and then controlled by respective
parametric solutions. Another important property is that such solutions can
be characterized by a G. Perelman thermodynamics generalized for
nonassociative Ricci flows \cite%
{perelman1,partner02,partner04,partner05,partner06} as we prove in the next
subsection.

\subsection{Examples of phase space fixed points and RG flows}

Let us consider two examples of fixed sets of values of $\left( \varrho _{\breve{\imath}}(\tau ), 
\ ^{\shortmid }\varrho _{\breve{\imath}}(\tau)\right) $ (\ref{dimenssc}) resulting in fixed points, for a fixed geometric flow parameter $\tau _{0}$ and analyse the respective behaviour of RG flows. In the G. Perelman approach, this is a temperature-like parameter when the explicit values  can be arbitrary ones depending on a conventional effective temperature scale (for instance, we can consider any $\tau _{0}:$
$0\leq \tau _{0}\leq 1).$  Such properties possess certain sub-classes of quasi-stationary off-diagonal solutions (\ref{ans1rf}) with nonlinear symmetries (\ref{nonlinsymr}) resulting in effective $\tau $-running
cosmological constants $\lambda (\tau ):=\ _{\check{p}}\Lambda (\tau )\check{%
p}^{-2}$ and $\ ^{\shortmid }\lambda (\tau ):=\ _{\check{p}}^{\shortmid
}\Lambda (\tau )\check{p}^{-2},$ respectively, for $_{\ast }\lambda =\lambda
(\tau _{0})$ and $\ _{\ast }^{\shortmid }\lambda =\ ^{\shortmid }\lambda
(\tau ).$ Extending on nonassociative phase spaces the Wilsonian viewpoint on cotangent Lorentz bundles 
$\ ^{\shortmid }\mathcal{M}^{\star }$ with parametric nonassociative star product deformations, we connect the issue of renormalizability to some fixed points $(\ _{\ast }\varrho _{\breve{\imath}%
},\ _{\ast }^{\shortmid }\varrho _{\breve{\imath}}),$ corresponding to an underlying nonassociative Ricci soliton and respective RG flow. Here we note that the RG flows are not identic to the geometric flows which involve more general constructions  may result in non-renormalizable solutions. From the viewpoint of RG flows, the fixed points are defined by the conditions $(\beta _{\breve{\imath}}, 
\ ^{\shortmid }\beta _{\breve{\imath}})_{\mid \ _{\ast }\varrho _{\breve{\imath}}, 
\ _{\ast }^{\shortmid }\varrho_{\breve{\imath}}}=0$ (in this subsection, we dub on base spacetimes and
cofibers, and then subject to nonholonomic star product deformation the constructions from \cite{gies16}. Linearizing such nonassociative phase space beta-functions at a fixed point, we can construct the stability
coefficients $(\theta _{\breve{\imath}},\ ^{\shortmid }\theta _{\breve{\imath}})$ defined for respective couples of base and cofiber matrices, $\left( B_{\breve{\imath}\breve{j}}:=\partial _{\varrho _{\breve{\imath}}}\beta _{\breve{\imath}\mid \ _{\ast }\varrho _{\breve{\imath}}},\ ^{\shortmid }B_{%
\breve{\imath}\breve{j}}:=\partial _{\ ^{\shortmid }\varrho _{\breve{\imath}}}\ ^{\shortmid }\beta _{\breve{\imath}\mid \ _{\ast }^{\shortmid }\varrho _{\breve{\imath}},}\right) .$ These define the free parameters of the nonassociative phase theory, which are associated with the stability coefficients (i.e. the corresponding positive real parts) to be determined experimentally.

We can choose the horizontal data and effective cosmological constants, and dub them as covertical ones for a class of solutions (\ref{ans1rf}) when the system of beta functions (\ref{betah}) is computed for LC-configurations and has a behaviour similar to that stated in  \cite{reuter98,souma99,reuter02,gies16}, i.e.%
\begin{equation}
\begin{array}{ccccc}
^{LC}GEP: & _{\ast }\lambda =0, & _{\ast }\varrho =0; & \ _{\ast
}^{\shortmid }\lambda =0, & \ _{\ast }^{\shortmid }\varrho =0. \\ 
^{LC}NGEP: & _{\ast }\lambda =0.193 & _{\ast }\varrho =0.707; & \ _{\ast
}^{\shortmid }\lambda =0.193 & \ _{\ast }^{\shortmid }\varrho =0.707.%
\end{array}
\label{e14}
\end{equation}%
Such primary configurations with respective asymptotic behaviour on the base Lorentz manifold can always be  generated as in GR by respective $\ _{s}^{\shortmid }\mathbf{\mathring{g}}$ chosen for (\ref{offdiagdefr}).

The solutions with $^{LC}GEP$ as in  (\ref{e14}) define free models of nonassociative gravity possessing a saddle point for which trajectories with a positive Newton constant in the h-part do not end at high energies. This reflects star product deformations of the off-diagonal solutions for the effective Einstein-Hilbert action. Such data can be selected by prescribing generating and integrating data (\ref{integrfunctrf}) to result in 
$\ ^{\shortmid }\widehat{\mathcal{T}}_{\ \star \beta \gamma }^{\alpha }=0,$ see formulas (\ref{twoconsstar}), and the h-metrics result in horizontal beta functions with properties from formulas (14) in \cite{gies16}. The formulas are dubbed on co-fibers by choosing (\ref{integrfunctrf}) to determine phase
space solutions with nontrivial nonholonomic canonical d-torsion in a form when the conditions $_{\ast }\lambda =\ _{\ast }^{\shortmid }\lambda =0$ are effective ones when $\ _{s}^{\shortmid }\Lambda (\tau _{0})$ can be nonzero.

The $^{LC}NGEP$ from (\ref{e14}) exhibits a complex pair of stability coefficients $\theta _{1,2}=1.475\pm 3.043i,$ which means that such configurations are UV attractive for both the Newton constant and the
effective constant when the asymptotic safety is guaranteed for the LC-conditions. Such a property can be preserved at least for small parametric off-diagonal decompositions for canonical s-configurations and
even under $\tau $-parametric flows of solutions preserving the same behaviour for $^{LC}NGEP\rightarrow NGEP$ with some values $(\ _{\ast }\lambda (\tau ), \ _{\ast }\varrho (\tau ); \ _{\ast }^{\shortmid }\lambda (\tau ), \ _{\ast }^{\shortmid }\varrho (\tau ))$ closed to (\ref{e14}).

It should be noted that the beta functions (\ref{betasigma}) clarifies the properties of the fixed point structure (\ref{e14}) even the term Goroff-Sagnotti phase space term $\Gamma _{\check{p}}(\tau )=
 \ _{\check{p}}^{\shortmid }\widehat{\mathcal{W}}_{\kappa }^{\star }(\tau )+
 \ _{\check{p}}^{\shortmid }\widehat{\Gamma }_{\kappa }^{GS}(\tau )$ (\ref{effectact}) is taken into account. Introducing zero $\lambda $ and $\varrho $ values into the formulas for beta functions, we can map and dub respectively%
\begin{equation}
\begin{array}{ccccccc}
^{GS}GEP: & _{\ast }\lambda =0, & _{\ast }\varrho =0, & _{\ast }\sigma =-%
\frac{7}{128\pi ^{2}}; & \ _{\ast }^{\shortmid }\lambda =0, & \ _{\ast
}^{\shortmid }\varrho =0, & \ \ _{\ast }^{\shortmid }\sigma =-\frac{7}{%
128\pi ^{2}}.%
\end{array}
\label{e15}
\end{equation}%
These formulas dub on $\ ^{\shortmid }\mathcal{M}$ and then extend on parametric nonassociative spaces 
$\ ^{\shortmid }\mathcal{M}^{\star }$ the results of formulas (16) from \cite{gies16} (to transform such solutions into EH+GS configurations from that work, we have to select solutions (\ref{ans1rf})  with trivial cofiber parts).

The priority of the AFCDM method is that it allows us to construct very general classes of off-diagonal solutions for nonassociative geometric flow and Ricci solition equations determined generating data (\ref{integrfunctrf}) and nonlinear symmetries (\ref{nonlinsymr}). We can always select certain primary s-metric configurations with fixed points of type (14)-(18) and Figure 1 from \cite{gies16} which allow us to state at the very beginning certain fixed points and clarify the conditions of the asymptotic safety on
the h-part of s-metrics. Such conditions can also be dubbed on cofibers which guarantee the existence of asymptotic safe quasi-stationary configurations determined by parametric nonassociative R-flux deformations.
In a more general context, when the primary data for s-metrics 
$\ _{s}^{\shortmid}\mathbf{\mathring{g}}(\tau )$  (see formulas (\ref{offdiagdefr})) are different from those including configurations investigated in \cite{reuter98,souma99,groh11,reuter02,gies16}, we may to apply the
Mathematical package xAct, 899 Goroff-Sagnotti terms corresponding to various vertexes etc. as mentioned in footnote \ref{fncompt}. In this work, we do not use computer methods but our nonassociative geometric formalism and the AFCDM allows us to state nonassociative configurations possessing asymptotic
safety in the h-part and then to study their off-diagonal and $\tau $-evolution deformations with properties of type (\ref{e14}) \ and (\ref{e15}).

\subsection{Thermodynamic variables for nonassociative RG and geometric flows%
}

Any nonassociative off-diagonal parametric geometric flow solution described
by a quasi-stationary configuration (\ref{ans1rf}) with coefficients (\ref%
{qstcoef}) and nonlinear symmetries (\ref{nonlinsymr}) is characterized by
generalized G. Perelman variables (see explicit examples and computations in 
\cite{partner04,partner05}). The goal of this subsection is to prove that
the classical nonassociative geometric flow evolution can be re-defined in
quantum variables for a phase space model of asymptotic safe thermo-field
theory and $\tau $-running families of beta functions (\ref{betah}) and (\ref%
{betasigma}).

We introduce the dimensionless cosmological constants $\lambda (\tau ):=\ _{%
\check{p}}\Lambda (\tau )\check{p}^{-2}$ and $\ ^{\shortmid }\lambda
(\tau):=\ _{\check{p}}^{\shortmid }\Lambda (\tau )\check{p}^{-2}$ from (\ref%
{dimenssc}) into formulas (89) and (90) from \cite{partner06} for the class
of $\tau $-running quasi-stationary s-metrics. This allows us to compute the
respective canonical statistical distribution function, $\ _{\check{p}%
}^{\shortmid }\widehat{\mathcal{Z}}_{\kappa }^{\star }(\tau ),$ statistical
and nonassociative geometric thermodynamic energy and entropy, $\ _{\check{p}%
}^{\shortmid }\widehat{\mathcal{E}}_{\kappa }^{\star }(\tau )$ and $\ _{%
\check{p}}^{\shortmid }\widehat{\mathcal{S}}_{\kappa }^{\star }(\tau)= -\ _{%
\check{p}}^{\shortmid}\widehat{\mathcal{W}}_{\kappa }^{\star }(\tau) $ (as
minus W-entropy considered in (\ref{naffunctpfh}) and (\ref{effectact})): 
\begin{eqnarray}
\ _{\check{p}}^{\shortmid }\widehat{\mathcal{Z}}_{\kappa }^{\star }(\tau )
&=&\exp [ \int\nolimits_{\tau ^{\prime }}^{\tau }\frac{d\tau }{(2\pi \tau 
\check{p})^{4}}\frac{1}{|\lambda (\tau )\ \ ^{\shortmid }\lambda (\tau )|}\
\ \ _{\check{p}}^{\shortmid }\mathcal{\bar{V}}(\tau )] ,  \notag \\
\ _{\check{p}}^{\shortmid }\widehat{\mathcal{E}}_{\kappa }^{\star }(\tau )
&=& -\int\nolimits_{\tau ^{\prime }}^{\tau }\frac{d\tau }{128 (\pi )^{4}\tau
^{3}}\frac{\tau \check{p}^{2}[\lambda (\tau )+\ ^{\shortmid }\lambda (\tau
)]-2}{\check{p}^{4}|\lambda (\tau )\ \ ^{\shortmid }\lambda (\tau )|} \ _{%
\check{p}}^{\shortmid }\mathcal{\bar{V}}_{\kappa }(\tau ),  \notag \\
\ \ _{\check{p}}^{\shortmid }\widehat{\mathcal{S}}_{\kappa }^{\star }(\tau )
&=&-\ _{\check{p}}^{\shortmid }\widehat{\mathcal{W}}_{\kappa }^{\star }(\tau
)=-\int\nolimits_{\tau ^{\prime }}^{\tau }\frac{d\tau }{(4\pi \tau \check{p}%
)^{4}}\frac{2(\tau \check{p}^{2}[\lambda (\tau )+\ ^{\shortmid }\lambda
(\tau )]-4)}{|\lambda (\tau )\ \ ^{\shortmid }\lambda (\tau )|}\ \ \ _{%
\check{p}}^{\shortmid }\mathcal{\bar{V}}_{\kappa }(\tau ).
\label{statistper}
\end{eqnarray}%
In formulas (\ref{statistper}), the volume functionals with momentum cutting
in phase spaces are defined and computed as $\ _{\check{p}}^{\shortmid }%
\mathcal{\bar{V}}_{\kappa }(\tau )= \int_{\ _{s}^{\shortmid }\widehat{\Xi }%
}\ ^{\shortmid }\delta \ _{\check{p}}^{\shortmid }\mathcal{\bar{V}}_{\kappa
}(\ ^{\shortmid }\eta _{\alpha _{s}} \ ^{\shortmid }\mathring{g}%
_{\alpha_{s}};\ _{s}^{\shortmid }\widehat{\Im }^{\star })$ using also
nonlinear symmetries (\ref{nonlinsymr}). In explicit form, such running
phase space volume functionals can be computed if we prescribe certain
classes of generating $\eta $-functions, effective generating sources $\
_{s}^{\shortmid }\widehat{\Im }^{\star },$ when the coefficients of a prime
s-metric $\ ^{\shortmid }\mathring{g}_{\alpha _{s}}$ and nonholonomic
distributions are defined as a closed hyper-surface $\ _{s}^{\shortmid }%
\widehat{\Xi }\subset \ _{s}^{\shortmid }\mathcal{M}^{\star }$ in vicinity
of a $\check{p}.$ Such computations can be performed for nonassociative BH,
WH and locally anisotropic solutions as in \cite%
{partner04,partner05,partner06}.

For small parametric deformations, the formulas (\ref{statistper}) describe
star product deformed physical objects under nonassociative geometric flows
when quantum fluctuations are nonholonomically constrained in some forms
which result in asymptotic safety models both on total phase space and for
projections of Lorentz spacetime manifolds. Such formulas relate classical
nonassociative geometric flows to beta functions (\ref{betah}) and (\ref%
{betasigma}) and their cofiber analogues which are important for
constructing physically viable models for nonassociative phase space QG.

\section{Conclusions}

\label{sec05}We have studied the modified gravity and nonassociative
geometric flow theory defined by star product R-flux deformations from
string theory. Nonassociative structures settle a long-standing question if
quantum gravity, QG, models constructed on phase spaces (cotangent Lorentz
bundles) may preserve the asymptotic safety property which exists for
quantizing the Einstein-Hilbert action supplemented by the two-loop
conterterm found by Goroff and Sagnotti \cite{goroff1,goroff2,ven92}.
Nonassociativity radically modifies general relativity and other types of
modified gravity theories resulting in quasi-classical parametrical limits
in new types of gravitational and (effective) matter field interactions and
geometric evolution scenarios on phase spaces. In such nonassociative phase
space theories, the fundamental geometric objects depend on conventional
spacetime and momentum, or velocity, like variables, when generic
off-diagonal metrics (with induced nonsymmetric terms) and effective
generating sources encode nonassociative data and subjected to respective
nonlinear symmetries. In a series of recent works \cite%
{blumenhagen16,aschieri17,partner04,partner05,partner06}, self-consistent
and physically viable nonassociative theories with modified black hole,
wormhole, locally anisotropic cosmological solutions and respective
geometric and quantum information flow theories were elaborated, when
classical real solutions were constructed for parametric decompositions on
the Planck and string constants, $\hbar $ and $\kappa .$

Using star product deformations, all versions of beta functions can be
defined and computed on nonassociative phase spaces in a form including the
Eintein-Hilbert approximation on base Lorentz manifold and canonical
nonholonomic distortions to the total Lorentz cotangent bundle. In general,
nonlinear symmetries relate nontrivial effective sources and generic
off-diagonal interactions to certain effective dyadic shell cosmological
constants. Nonassociative configurations may evolve as geometric flows on an
effective temperature like parameter $\tau $ being characterized by a
corresponding generalized G. Perelman thermodynamics \cite%
{perelman1,partner04,partner06}. Such nonassociative geometric methods
generalize the constructions on renormalization group flows by adding new
types of geometric thermodynamic variables associated with new classes of
off-diagonal solutions which, in general, destroy the non-Gaussian fixed
points seen in the projections to Einstein-Hilbert configurations.

Nevertheless, for corresponding classes of off-diagonal solutions and
nonlinear symmetries (considering, for instance, at the end some effective
zero/ small cosmological limits) we may select nonholonomic configurations
on the base spacetime manifold (including the full feedback of the
counter-terms), which posses a non-Gaussian fixed point in agreement with
the gravitational asymptotic safety scenario. To simplify the constructions,
we can consider certain classes of off-diagonal metrics which are effective
conformally flat on the total phase space (with zero Weyl tensor for the
canonical s-connection) but contain nontrivial terms for projections on
Lorentz manifold base. The constructions allow selection of solutions
describing trajectories with a (semi-) classical low-energy regime remaining
un-tach. So, nonassociativity, at least for small parametric deformations
and corresponding G. Perelman thermodynamic functionals, with verification
of locality and selection of generation functions to keep unitarity, allows
satisfying the conditions of the asymptotic safety program. This means that
the class of nonassociative QG theories involving star products with
R-fluxes from string theory are viable as QFTs in the lines stated by
Weinberg and further constructions \cite%
{weinberg79,niedermaier06,gies16,bonanno11,dietz13,benedetti13,ohta15}.

\vskip5pt \textbf{Acknowledgement:}\ The author thanks Prof. Douglas
Alexander Singleton for hosting a Fulbright visiting research program in the
USA, on nonassociative geometric and information flows; Prof. Dieter L\"{u}%
st for hosting a scientist at risk program at LMU, Munich, Germany. Both
programs resulted in this sub-program on nonassociative QG theories for a
volunteer research in Ukraine hosted by Prof. Valery I. Zhdanov. He is also
grateful for arranging new collaborations in Ukraine to Prof. Julia O. Seti
and Prof. Alexander Zhuk and for support and long term collaboration by
Prof. Panayotis Stavrinos, Greece; Prof. El\c{s}en V. Veliev, Turkey; and
Dr. Lauren\c{t}iu F. Bubuianu, Romania.

\end{document}